\documentclass[onecolumn,preprintnumbers,amsmath,amssymb, longbibliography]{revtex4}
\usepackage{graphicx}
\usepackage{dcolumn}
\usepackage{bm}
\usepackage{epsfig}
\usepackage{subfigure}
\usepackage{xcolor}
\usepackage{amsthm}
\usepackage{mathtools}

\newcommand{\bi}{\begin{itemize}}
\newcommand{\ei}{\end{itemize}}
\newcommand{\be}{\begin{eqnarray}}
\newcommand{\ee}{\end{eqnarray}}
\newcommand{\beq}{\begin{equation}}
\newcommand{\eeq}{\end{equation}}
\newcommand{\beqn}{\begin{equation*}}
\newcommand{\eeqn}{\end{equation*}}
\newcommand{\dd}{\text{d}} 
 
\newcommand{\bbmatrix}{\left( \begin{array}}
\newcommand{\eematrix}{\end{array} \right)}

\begin{document}

\title{Asymptotic behavior of continuous weak measurement and its application to real-time parameter estimation}

\author{Chungwei Lin$^1$\footnote{clin@merl.com}, Yanting Ma$^1$} %, Yebin Wang$^1$ }
\affiliation{$^1$Mitsubishi Electric Research Laboratories, 201 Broadway, Cambridge, MA 02139, USA 
} 

\author{Dries Sels$^{2,3}$}
\affiliation{ $^2$Department of physics, New York University, New York City, NY 10003, USA \\
$^3$Center for Computational Quantum Physics, Flatiron Institute, 162 5th Ave, New York, NY 10010, USA\footnote{The Flatiron Institute is a division of the Simons Foundation.} }

%\author{Chungwei Lin$^1$\footnote{clin@merl.com}, Dries Sels$^{2,3}$, Yebin Wang$^1$}
\date{\today}

\begin{abstract} 
The asymptotic quantum trajectory of weak continuous measurement for the magnetometer is investigated. The magnetometer refers to a setup where the field-to-estimate and the measured moment are orthogonal, and the quantum state is governed by the stochastic master equation which, in addition to a deterministic part, depends on the measurement outcomes.
We find that the asymptotic behavior is insensitive to the initial state in the following sense: given one realization, the quantum trajectories starting from arbitrary initial states asymptotically converge to the {\em same} realization-specific {\em pure} state.
For single-qubit systems, we are able to prove this statement within the framework of Probability Theory by deriving and analyzing an effective one-dimensional stochastic equation.
Numerical simulations strongly indicate that the same statement holds for multi-qubit systems.
%Our results imply that when a quantum task involves the continuous measurement, the initial state is not and cannot be critical asymptotically.
Built upon this conclusion, we consider the problem of real-time parameter estimation whose feasibility hinges on the insensitivity to the initial state, and explicitly propose and test a scheme where the quantum state and the field-to-estimate are updated simultaneously.
\end{abstract}

%\pacs{31.15.A-,71.55.-i,73.20.hb}
\maketitle

%%%%%%%%%%%%%%%%%%%%%%%%%%%%%%%%%%%%%%%%%%% 

\section{Introduction}  

Wave function collapse upon measurements is one of the quantum features that has no classical analog. Contrary to the projective measurement where the system is in the observable eigenstate immediately after the measurement, the weak measurement refers to the scenario where an observer gains very little information and at the same time the system is only infinitesimally disturbed from a single experimental outcome.
The quantum state evolution conditioned on the outcomes of continuous weak measurements, also known as the quantum trajectory, is governed by the stochastic master equation (SME) that includes not only the deterministic Lindblad dynamics but also the stochastic ``measurement back-action''.
%In other words, a quantum trajectory not only depends on the Hamiltonian but a particular sequence of measurement-specific outcomes. 
The quantum trajectory theory \cite{book_Wiseman_Milburn, book:Carmichael, book:Barchielli, book:Jacobs, Belavkin_1992,  GISIN1992315, 10.1119/1.1475328, PhysRevA.49.2133, Wiseman_1996_QuantumTrajectory, Jacobs-2006} not only deepens our understanding of measurement process but also leads to many theoretical and experimental developments.
Averaging quantum trajectories over potential experimental outcomes can be used as a memory efficient numerical scheme to simulate the open quantum system \cite{PhysRevA.45.4879, Molmer_93, book:open_quantum}. The quantum trajectory conditioned on the measurement outcomes has been experimentally realized in superconducting qubits \cite{doi:10.1126/science.1226897, Nature.502.211_Murch_2013, PhysRevLett.114.090403, WEBER2016766}.
For applications, the continuous measurement provides perhaps an easier and more robust route to entangle a large number of small quantum objects \cite{Sorensen_2001, PhysRevA.89.043837, PhysRevLett.85.1594, PhysRevLett.93.173002, Hosten_2016}.  It has also been formulated for quantum metrology where the measurement outcomes are not only used to update the quantum trajectory but also to estimate the parameter(s) in the Hamiltonian \cite{PhysRevLett.91.250801, PhysRevLett.93.173002, PhysRevA.87.032115}. It is shown that  combining the continuous weak measurement and one project measurement can in principle enhance the sensing performance against certain quantum decoherence channels \cite{Albarelli_2017, Albarelli2018restoringheisenberg, PhysRevLett.125.200505}. Moreover, measurement outcomes can be used as feedbacks to actively control the quantum system to achieve certain states \cite{PhysRevA.70.022106, PhysRevLett.75.4587, PhysRevLett.85.5098, Pozza_2015, Hacohen.Gourgy.2016, Martin_2020}, to accelerate the purification \cite{PhysRevA.63.062305, PhysRevA.67.030301, PhysRevLett.96.010504, Wiseman_2006,  PhysRevA.82.022307, Ruskov_2012}, or for quantum state estimation \cite{PhysRevLett.115.180407, PhysRevA.91.012118, PhysRevA.96.052306, PhysRevA.104.052621}.

In this work we investigate the asymptotic behavior of a quantum trajectory under the continuous weak measurement for magnetometer. The magnetometer refers to a setup where the external field and the measured moment are orthogonal, and one estimates the field strength based on the experimental outcomes \cite{PhysRevLett.91.250801, Albarelli_2017, Albarelli2018restoringheisenberg, PhysRevLett.125.200505}. For weak measurements the quantum trajectory is governed by the stochastic master equation (SME) and is formally a stochastic process \cite{Book_SP_Evans, Book_SP_Allen, Book_SP_Arnold, Book_SP_Khasminskii, book_Shreve_I, book_Shreve_II}, and we apply well developed concepts/tools in Stochastic Differential Equation (SDE) and Probability Theory to analyze our results.
In order for an analytical analysis we do not include quantum decoherence or feedback controls.
%To analyze the results, we formulate the SME in the framework of Stochastic Process \cite{Book_SP_Evans, Book_SP_Allen, Book_SP_Arnold, Book_SP_Khasminskii, book_Shreve_I, book_Shreve_II}; well developed concepts/tools in Stochastic Differential Equation (SDE) and Probability Theory %,  particularly ``convergence in probability'' and Fokker-Planck (FP) equation (Chapman-Kolmogorov equation)  are used to quantitatively analyze the numerical observations.
Our results are consistent with the following statement: the asymptotic behavior depends only on the measurement back-action, not on the initial state at all. In the framework of SDE,
each set of measurement outcomes is adapted to a realization of Brownian motion, and all quantum trajectories adapted to the same realization  are evolved to the {\em same} time-varying realization-dependent {\em pure} state asymptotically. Although this statement could be anticipated from some general arguments  (see Chapter 3 of Ref.~\cite{book:Jacobs} and Ref.~\cite{doi:10.1142/S0219025709003549}), a general proof is still lacking. We are able to prove this statement for single-qubit systems, but numerical simulations strongly support its validity for multi-qubit systems and also generic Hamiltonians.
Built upon this conclusion we consider the problem of real-time parameter estimation whose feasibility replies on the insensitivity to the initial state, and explicitly devise and test an SDE that simultaneously estimates the full quantum state and the parameter.

Our proof is divided into two steps: (i) the asymptotic state is pure; (ii) given a realization any two initial pure states eventually coincide.
Step (i), the asymptotic purity, may not be surprising based on the conclusions of Quantum-Nondemolition (QND) measurement \cite{PhysRevA.60.2346, PhysRevLett.85.1594} and some purification feedback control schemes \cite{van_Handel_2005, Benoist-2014, 8618681, PhysRevA.63.062305, PhysRevA.67.030301, PhysRevLett.96.010504, Wiseman_2006,  PhysRevLett.100.160503, PhysRevA.82.022307, Ruskov_2012}. Here we consider the non-QND measurement without controls. The proof involves the concept of ``convergence in probability'' and a few well-known inequalities. Step (ii) is perhaps less obvious, and Jacobs in Ref.~\cite{book:Jacobs} provides an argument by regarding infinitely many weak measurements as a projection operator to a pure state. Our proof involves solving the time-independent Fokker-Planck (FP) equation (Chapman-Kolmogorov equation) using the ``continued fraction'' technique \cite{book_SP_Gardiner, book_Langevin_Coffey, book_FP_Risken}. %Although the problem considered here is specific, we believe it can be a useful test bed for many interesting physics such as quantum decoherence and measurement efficiency.
As for the real-time parameter estimation scheme, the most crucial step is to derive the SDE for the gradient of the log-likelihood function \cite{PhysRevA.87.032115}.

%Insensitivity to the initial states suggests the possibility of real-time parameter estimation; we demonstrate it by devising a SDE that simultaneously estimates the full quantum state and the parameter.
%Our analysis strongly implies that when a quantum task involves the continuous measurement, the initial state is not sensitive to the asymptotical state. As a test to our conclusion and also a potential application,

%It has been established that the system eventually collapses to one of the measurement eigenstates under the Quantum-Nondemolition (QND) measurement \cite{PhysRevA.60.2346, PhysRevLett.85.1594}. With feedback control one can guide the system to a specific measurement eigenstate \cite{van_Handel_2005, Benoist-2014, 8618681} or speedup the purification process \cite{PhysRevA.63.062305, PhysRevA.67.030301, PhysRevLett.96.010504, Wiseman_2006,  PhysRevLett.100.160503, PhysRevA.82.022307, Ruskov_2012}. One key difference of the magnetometer setup is that the measurement is generally {\em not} QND and the asymptotic pure state keeps on evolving in time.

The paper is organized as follows. In Section II we present the concrete problem of interest, derive the equation for single-qubit system and show the numerical evidences that support our statement. 
In Section III we collect a few mathematical tools and frameworks needed for the analysis. 
In Section IV the probability theory is used to show the asymptotic purity. 
In Section V the FP equation for single qubit is derived and then solved by the technique of continued fraction. Using the stationary distribution we conclude that that any two quantum trajectories eventually coincide asymptotically. Experimental implications, including a non-rigorous generalization, are discussed; comparisons to existing works are provided.
In Section VI we develop a real-time parameter estimation scheme based on local maximum likelihood, whose feasibility is motivated by the conclusion of Section V.
A brief conclusion is given in Section VII. Appendixes provide important intermediate steps skipped in the main text. In Appendix \ref{app:stationary} we discuss the asymptotic and stationary behavior of FP equation. In Appendix \ref{app:CF}  some details about continued fractions are provided. In Appendix \ref{app:Lyapunov} we apply the Lyapunov analysis to a special case. In Appendix \ref{app:SDE_LL} we derive the SDE for the log-likelihood function and its derivatives.

%%%%%%%%%%%%%%%%%%%%%%%%%%%%%%%%%%

\section{System of consideration and numerical evidence}

\subsection{Overview}

The SME for the density matrix (DM) for the magnetometer setup is
\begin{subequations} 
\begin{align}
\dd \rho_t = &  -i \big[(- B \hat{J}_y), \rho_t \big] \dd t + \mathcal{D}[\hat{J}_z] \rho_t \, \dd t \nonumber \\
& + \left\{ \rho_t \hat{J}_z + \hat{J}_z \rho_t -  2 \ \text{tr}[ \rho_t \hat{J}_z]   \rho_t \right\} \cdot \left\{ \dd Y_t - 2 \ \text{tr}[ \rho_t \hat{J}_z]  \dd t  \right\},  
\label{eqn:SDE_general} \\
%%%%%%%%%%%%%55
\dd Y_t &=2 \ \text{tr}[ \rho_t \hat{J}_z] \ \dd t  + \dd W_t.
\label{eqn:SDE_outcome}
\end{align}
\label{eqn:drho_SDE}
\end{subequations} 
In Eq.~\eqref{eqn:SDE_general}, $\hat{J}_\alpha = \frac{1}{2} \sum_{i=1}^N \sigma_{i,\alpha}$ and $\mathcal{D}[\hat{J}_z] \rho  = -\frac{1}{2} \{ \hat{J}_z^2, \rho \} + \hat{J}_z \rho \hat{J}_z^\dagger $.  In Eq.~\eqref{eqn:SDE_outcome}, $\dd Y_t$ is the measurement outcomes and $\dd W_t$ is the Wiener increment which is stochastic and has a normal distribution of zero mean and $\sqrt{\dd t}$ standard deviation, i.e., $\dd W_t \sim \mathcal{N}(0, \sigma^2 = \dd t)$. $W_t$ is referred to as the Brownian motion or Wiener process (see Chapter 3 of Ref.~\cite{Book_SP_Evans}).
Eq.~\eqref{eqn:drho_SDE} models the quantum system of $N$ atoms in a magnetic field of  $-B \hat{y}$ (the minus sign is chosen for convenience) whose overall magnetization along a orthogonal direction $z$ ($\sim$$\langle \hat{J}_z \rangle$) is continuously monitored. %; we shall refer the conditional evolution of Eq.~\eqref{eqn:SDE_general} as a ``magnetometer'' setup \cite{PhysRevLett.91.250801, Albarelli_2017}.
It is worth mentioning that Eq.~\eqref{eqn:SDE_general} can be understood as a quantum state estimator \cite{PhysRevA.69.032109}. In this context the deterministic part is the ``model prediction'' which is the best state estimation without measurements; the stochastic part (term in second line) is the ``innovation'' that corrects the model prediction from the measurement outcome $\dd Y_t$.

When $B=0$, the system Hamiltonian (which is zero here) commutes with the measurement operator $\hat{J}_z$. This is known as the QND measurement \cite{PhysRevA.60.2346, PhysRevLett.85.1594}  and the system will asymptotically collapse one of the eigenstates of measurement operator.  In this work we focus on the asymptotic behavior of $B \neq 0$.  %The main statements  will be given in Section~\ref{sec:main_finding}. %As the analytical analysis is done only for the single-qubit systems,
%Although based on simulations the statements remain general for system of many qubits, the analytical analysis is done only for the single-qubit system.

\subsection{Simulation procedure and equation for single qubit}

For a simulation, we discretize the total evolution time $T$ into $N$ equally-spaced intervals $[0, \dd t, 2 \dd t, \cdots, N \dd t]$ (i.e., $t_n = n \ \dd t$ and $T= N \dd t$) and generate a list of mutually independent Wiener increments $[\dd W_1, \dd W_2, \cdots, \dd W_N]$ according to $\mathcal{N}(0, \sigma^2=\dd t)$. Eqs.~\eqref{eqn:drho_SDE} %[also Eq.~\eqref{eqn:SME_1Qubit_comp} and \eqref{eqn:dY_1Qubit_comp}]
are simulated by
\begin{subequations}
 \begin{align}
 \dd \rho_{ (n+1) \dd t } &=  \rho_{ (n+1) \dd t } - \rho_{ n \cdot \dd t } \nonumber \\
 &= i \big[ B \hat{J}_y, \rho_{ n \cdot \dd t } \big] \dd t + \mathcal{D}[\hat{J}_z] \rho_{ n \cdot \dd t } \, \dd t \nonumber \\
& + \left\{ \rho_{ n \cdot \dd t } \hat{J}_z + \hat{J}_z \rho_{ n \cdot \dd t } -  2 \ \text{tr}[ \rho_{ n \cdot \dd t } \hat{J}_z]   \rho_{ n \cdot \dd t } \right\}  \dd W_{ (n+1) \dd t }
\label{eqn:SME_dis} \\
%%%%%%%%%%%%%%%
\dd Y_{ (n+1) \dd t } &=2 \ \text{tr}[ \rho_{ n \cdot \dd t } \hat{J}_z] \ \dd t  + \dd W_{ (n+1) \dd t }.
\label{eqn:dY_dis}
 \end{align}
\label{eqn:dY_discrete}
\end{subequations}
Each list of Wiener increments, denoted by $\{ \dd W_t \}$, is referred to as a {\em realization} of the Brownian motion or Wiener process, and the corresponding $\rho_t$ is the quantum trajectory {\em adapted} to the realization $\{ \dd W_t \}$. $\{ \dd Y_t \}_{\rho_\text{init}}$ represents a list of measurement outcomes generated from the initial state $\rho_\text{init}$. Experimentally it is $\dd Y_t$, not $\dd W_t$, that is directly measured. In practice, one uses the experimental outcome $\dd Y_{(n+1) \dd t}$ and Eq.~\eqref{eqn:dY_dis} to get $\dd W_{(n+1) \dd t}$, and then uses  Eq.~\eqref{eqn:SME_dis} to evolve $\rho_t$.
Determining a quantum trajectory between $[0, T]$ requires an initial state and either $\{ \dd W_t \}$ or $\{ \dd Y_t \}$ for $0 < t \leq T$.
We remark that as an integral equation Eq.~\eqref{eqn:SME_dis} obeys the It\^{o}'s integral rule where $\rho_{ (n+1) \dd t } $ depends only on $\rho_{ n \cdot \dd t }$, not on any information between $n \cdot \dd t$ and $(n+1) \cdot \dd t$.
To better preserve the trace and semi-positive definiteness during the evolution, Eq.~\eqref{eqn:SME_dis} is evaluated using the formalism based on Kraus operators given in Refs.~\cite{PhysRevA.91.012118, Albarelli2018restoringheisenberg}. %In this work the  analysis is done on the system of single qubit.

%Eq.~\eqref{eqn:drho_SDE} has
Our analytical analysis will be done only for single-qubit systems.
For a single qubit, Eq.~\eqref{eqn:drho_SDE} is reduced to  
\beq 
\begin{aligned}
\dd \rho_t =& i B\ [\frac{\sigma_y}{2}, \rho_t] \dd t +  \left( \frac{\sigma_z}{2} \rho_t \frac{\sigma_z}{2}  - \frac{1}{2} \big\{ \frac{\sigma_z^2}{4}, \rho_t \big\} \right)\dd t \\
&+  \left( \big\{ \frac{\sigma_z}{2}, \rho_t \big\} - 2 \ \text{tr}\big[\rho_t \frac{\sigma_z}{2} \big] \rho_t \right) \, \dd W_t . 
\end{aligned}
\label{eqn:SME_1Qubit}
\eeq 
Parametrizing a single qubit state as
$ 
\rho = \frac{1}{2} \big[ \mathbb{I} + \bm{\rho} \cdot \bm{\sigma} \big], \text{ with }
\bm{\rho} = [ \rho_x, \rho_y,  \rho_z ]^T
%\label{eqn:rho_real}
$, 
Eq.~\eqref{eqn:SME_1Qubit} in component form is
\beq 
\dd  \begin{bmatrix} \rho_x \\ \rho_y \\ \rho_z \end{bmatrix}
= B \begin{bmatrix} -\rho_z \\ 0 \\ \rho_x \end{bmatrix}  \dd t
- \frac{1}{2}  \begin{bmatrix} \rho_x \\ \rho_y \\ 0 \end{bmatrix} \dd t
+   \begin{bmatrix} -  \rho_z \rho_x \\ -  \rho_z \rho_y \\ 1-  \rho_z^2 \end{bmatrix} \dd W_t .
\label{eqn:SME_1Qubit_comp}
\eeq 
The measurement is given by 
\beq 
\dd Y_t = 2 \ \text{tr} \big[ \rho_t \frac{\sigma_z}{2} \big] \dd t + \dd W_t = \rho_z (t) \dd t+ \dd W_t. 
\label{eqn:dY_1Qubit_comp}
\eeq 
Semi-positive definiteness of DM requires $\rho_x^2 + \rho_y^2 + \rho_z^2 \leq 1$; the equality holds when the state is pure.

\subsection{System decoupling: polar coordinate} \label{sec:decoupling}

Eq.~\eqref{eqn:SME_1Qubit_comp} shows that $\rho_y$ does not affect $\rho_x$ or $\rho_z$, and if $\rho_y(0) = 0$ then $\rho_y(t)=0$ for all $t \geq 0$. In fact, we shall show $\rho(t) \rightarrow 0$ asymptotically for any $\rho_y(0)$ (Section~\ref{sec:asym_purity}), and for this reason we focus on the dynamics of $(\rho_x, \rho_z)$.
As the norm of $(\rho_x, \rho_z)$ indicates the purity of the DM, it is natural to use the polar coordinate $\rho_x + i \rho_z \equiv r e^{ i \theta}$ $\Leftrightarrow$ $r = \sqrt{\rho_x^2+\rho_z^2}$, $\theta = \tan^{-1}(\rho_z/\rho_x)$.  Using It\^{o}'s rule, the SDE's for $(r,\theta)$ are 
\beq 
\begin{aligned}
 \dd r &= \frac{\partial r}{ \partial \rho_x } \dd \rho_x + 
 \frac{\partial r}{ \partial \rho_z } \dd \rho_z + \frac{1}{2} 
 \bigg[ \frac{\partial^2 r}{ \partial \rho_x^2 } (\dd \rho_x)^2 +
 \frac{\partial^2 r}{ \partial \rho_z^2 } (\dd \rho_z)^2 
 + 2 \frac{\partial^2 r}{ \partial \rho_x \partial \rho_z } (\dd \rho_x  \dd \rho_z)
 \bigg], \\
 %%%%%%%%
 \dd \theta &= \frac{\partial \theta}{ \partial \rho_x } \dd \rho_x + 
 \frac{\partial \theta}{ \partial \rho_z } \dd \rho_z + \frac{1}{2} 
 \bigg[ \frac{\partial^2 \theta}{ \partial \rho_x^2 } (\dd \rho_x)^2 +
 \frac{\partial^2 \theta}{ \partial \rho_z^2 } (\dd \rho_z)^2 
 + 2 \frac{\partial^2 \theta}{ \partial \rho_x \partial \rho_z } (\dd \rho_x  \dd \rho_z)
 \bigg] .
\end{aligned} 
\nonumber
\eeq 
Substituting Eq.~\eqref{eqn:SME_1Qubit_comp} into the equations above gives 
\begin{subequations}
 \begin{align} 
  \dd r &=  -\frac{1}{2}  \cos^2 \theta \cdot (r - r^{-1}) \dd t + \sin \theta \cdot (1-r^2) \ \dd W_t , \label{eqn:dr_general} \\
  %%%%%%%%%
  \dd \theta &= \big( B + \frac{1}{4} \sin (2 \theta) \big) \dd t 
 +   \frac{1}{r} \cos \theta \cdot \dd W_t.
  \label{eqn:dtheta_general} 
 \end{align}
\label{eqn:SDE_polar_general}
\end{subequations}
For a pure state where $r=1$, $\dd r = 0$ implying a pure state remains pure and the corresponding SDE for the  angular variable is 
\beq 
\dd \theta  =  \big(B + \frac{1}{4} \sin (2 \theta) \big) \dd t  +  \cos \theta \cdot \dd W_t .
 \label{eqn:dtheta_pure}
\eeq 
When $B=0$, $\dd \theta=0$ when $\theta = \pm \frac{\pi}{2}$ which correspond to two eigenstates of measurement operator $\frac{\sigma_z}{2}$, and $\theta_t$ will be settled in either $\frac{\pi}{2}$ or $-\frac{\pi}{2}$ asymptotically. Once $B \neq 0$, there is no angles such that $\dd \theta=0$ so $\theta_t$ will keep on changing. % its valued according to a $B$-dependent stationary distribution.
Eq.~\eqref{eqn:dtheta_pure} is the 1D non-linear SDE that will be analyzed thoroughly in this work.

%We mention Eq.~\eqref{eqn:SDE_polar_general} remains valid if we replace $\dd W_t$ by \blue{ $\dd \tilde{W}_t = \dd W_t + \xi(t) \dd t$ } where  $\xi(t)$ is a stochastic process adapted to the same $\{ \dd W_t \}$.

\subsection{Main statement and illustrations} \label{sec:main_finding} 

\begin{figure}[ht]
\centering
\includegraphics[width=0.8\textwidth]{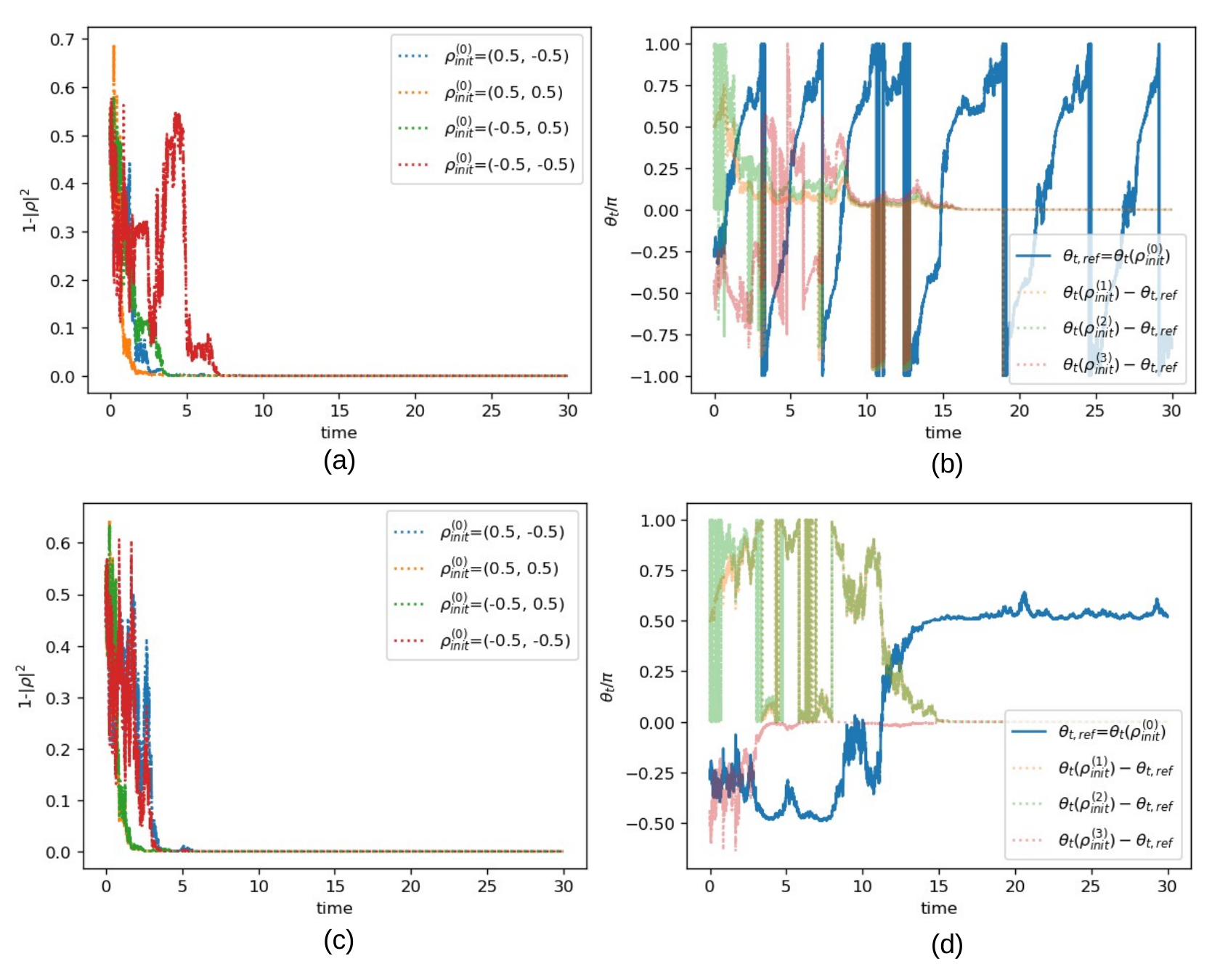}
\caption{ Illustration of the main result: given the same realization $\{ \dd W_t \}$ the system of different initial conditions evolve into the same $\{ \dd W_t \}$-specific pure state. Four initial states $\pmb{\rho}_\text{init} = (\rho_x, \rho_z)$ for  $B=1$ [(a), (b)] and $B=0.1$ [(c), (d)] are considered. The asymptotic purity is demonstrated in (a) and (c)  as $1 - |\pmb{\rho}|^2$ approaches zero. In (b) and (d) the angular variable $\theta_t$ defined by $\rho_x(t) + i \rho_z(t) \equiv e^{ i \theta_t}$ is plotted. The solid curve represents the reference $\theta_{t, \text{ref} }$ with $\pmb{\rho}_\text{init} = (0.5, -0.5)$. The convergence to the $\{ \dd W_t \}$-specific pure state is demonstrated in (b) and (d) where the difference $\theta_t ( \pmb{\rho}_\text{init} ) - \theta_{t, \text{ref} }$ approaches zero. 
}
\label{fig:1Qubit_main}
\end{figure} 

Our main statement can be summarized as follows: for a given realization $\{ \dd W_t \}$ and $B \neq 0$, the quantum trajectories starting from any initial states will asymptotically converge to the same realization-specific pure state.  
We shall present this statement in two steps. First we show that all quantum trajectories obeying Eq.~\eqref{eqn:SME_1Qubit_comp} become pure asymptotically; this will be referred to as ``asymptotic purity''.  Second we show that when $B \neq 0$ all initial state evolve into the same realization-specific pure state. 

Two numerical illustrations are given in Fig.~\ref{fig:1Qubit_main}. We consider four initial states from four quadrants in $\rho_x$-$\rho_z$ plane: $\pmb{\rho}^{(0)}_\text{init} = (\rho_x, \rho_z) = (0.5, -0.5)$, $\pmb{\rho}^{(1)}_\text{init} =  (0.5, 0.5)$, $\pmb{\rho}^{(2)}_\text{init} =  (-0.5, 0.5)$, $\pmb{\rho}^{(3)}_\text{init} =  (-0.5, -0.5)$ and evolve them for $B=1$ [Fig.~\ref{fig:1Qubit_main}(a), (b)] and $B=0.1$ [Fig.~\ref{fig:1Qubit_main}(c), (d)] using the {\em same} realization $\{ \dd W_t \}$. Note that these initial states are mixed states and $\rho_y$ is taken to be zero due to its irrelevancy as discussed in Section \ref{sec:decoupling}. Fig.~\ref{fig:1Qubit_main}(a) and (c) show that $1 - |\pmb{\rho}|^2$  becomes zero asymptotically for all simulations, meaning the state becomes pure. 
To show that all states converge to the same (pure) state, we solve $\pmb{\rho}$ with different initial conditions and obtain the corresponding angular variable $\theta_t (\pmb{\rho}_\text{init})$ (using $r_t e^{i \theta_t} = \rho_x(t) + i \rho_z(t)$). Fig.~\ref{fig:1Qubit_main}(b) and (d), the solid curve serves as the reference $\theta_{t, \text{ref}} = \theta_t (\pmb{\rho}^{(0)}_\text{init})$. The dashed curves are $\theta_t (\pmb{\rho}^{(i)}_\text{init}) - \theta_{t, \text{ref}}$; they all become zero asymptotically. %Generally the time for a quantum trajectory to become pure is shorter than the time for different quantum trajectories to become

So far the statement is based on observations of many numerical simulations. The next three sections are devoted to show the generality of these observations using the framework of probability theory and stochastic process.

%%%%%%%%%%%%%%%%%%%%%%%%%%%%
\section{Relevant mathematical formalisms}

Compared to the deterministic process, the randomness of the stochastic process makes a quantitative description less straightforward and it may not be immediately clear what it takes to prove a statement.
To facilitate the discussion, we collect a few relevant mathematical formalisms/tools in this section .

\subsection{Probability theory} 

For a continuous random variable $X$, its behavior is characterized by a probability distribution $P(X)$ or $P_X(x)$. For a random variable taking on continuous values $\mathbb{E}[X^n] = \int \dd x \ x^n P_X(x)$. A stochastic process is a collection of random variables, denoted as $X_t$ (or $X_n$), which is characterized by the joint probability $P(X_1, X_2, \cdots)$. The following mathematical  formalism/definition/theorem will be used in later discussion. 

\bi 

\item Markov inequality: For a positive random variable $X$, $P(X \geq a) \leq \frac{ \mathbb{E}[X] }{a}$.

\item Chebyshev Inequality: $P(|X| \geq \varepsilon ) \leq \frac{1}{\varepsilon^p} \mathbb{E}[|X|^p])$.

$p=1$ recovers the Markov inequality.
$p=2$ gives $P(|X| \geq \varepsilon ) \leq \frac{1}{\varepsilon^2} \mathbb{E}[X^2])$.

%Notice that  $\mathbb{E}[X^2]$ being unbounded does not necessarily mean $P(|X| \geq \varepsilon ) $ diverges; these are inequalities.

Two useful cases to bound $P(|X| \geq \varepsilon )$ are: (i) vanishing $ \mathbb{E}[|X|^p]$ with small but finite $\varepsilon^p$ and (ii) $ \mathbb{E}[|X|^p]$ diverges slower than  $\varepsilon^p$; both are used in Section \ref{sec:asym_purity}.

\item Convergence {\em in probability} \cite{Book_SP_Allen}. 
The sequence of random variables $\{ X_i \}_{i=1}^\infty$ is said to converge in probability to a fixed value $x$ if given any $\epsilon > 0$ there exists
a sufficiently large $n$ such that $P( |X_n - x| > \epsilon)< \delta$ for any small $\delta>0$; the latter statement is equivalent to $\underset{n \rightarrow \infty}{\lim} P (|X_n - x| > \epsilon) = 0$.

Convergence {\em almost surely} is a stronger statement than convergence in probability; we only consider the latter in this paper.

\item Change of variables in the probability density function (PDF): 
For a given PDF $P_X(x)$, the corresponding $P_Y(y)$ for the change of variable $y = f_t(x)$ is given by
\beq 
P_X (x) \dd x = P_Y (y) \dd y \Rightarrow 
P_Y(y) = \left( P_X(x) \big| \frac{\dd f_t}{\dd x} \big| \right)_{x = f_t^{-1}(y)}.
\label{eqn:P_variable_change}
\eeq  
Here $t$ is the parameter of the one-to-one transformation function $f_t$; in typical applications $t$ represents the time. Assuming $f_t$ is monotonously increasing, Eq.~\eqref{eqn:P_variable_change} implies that $P_X(x \geq x_0)= P_Y (y \geq f_t(x_0) )$.

\ei  

%As an example, $Y = F(X) \equiv e^X$ with $X \sim N(0, \sigma^2)$ or $P_X(x)  = \frac{1}{\sqrt{2 \pi} \sigma} e^{-x^2/(2 \sigma^2)}$. Using $x = \ln y$ and $\frac{\dd x}{\dd y} = \frac{1}{y}$, we get $P_Y(y)  = \frac{1}{\sqrt{2 \pi \sigma^2} y} e^{-(\ln y)^2/(2 \sigma^2)}$.

\subsection{Stochastic differential equation and stochastic process} 

%Some relevant statements are summarized for clarity. 

Consider a stochastic process $X_t$ that obeys the SDE 
\beq 
\dd X_t = f(X_t, t) \dd t + g(X_t, t) \dd W_t.
\nonumber 
\eeq 
$X_t$ is a Markov process as $X_{t + \dd t}$ depends only on $X_t$. 

\bi 

\item Time-homogeneous process is defined via the transition probability: 
\beq 
P(x_1, t_1 | x_0, t_0) = P(x_1, t_1 - t_0 | x_0, 0) \nonumber
\eeq 
This is true when $f$ and $g$ does not explicitly depend on $t$. The system of interest described by Eq.~\eqref{eqn:SDE_general} satisfies this property.

\item Stationary (strictly) process is defined via the joint probability: 
\beq 
P(x_0, t_0; x_1, t_1 ; \cdots; x_n, t_n ) = 
P(x_0, t_0+h; x_1, t_1+h ; \cdots; x_n, t_n+h )
\nonumber
\eeq 
For a stationary process, its statistics remains invariant under time translations. Stationary implies time-homogeneous, but the converse is not true.

\item 
(Definition) {\em Adapted:} A stochastic process $\Delta(t)$ is {\em adapted} to $X_t$ is for all $t \geq 0$ $\Delta(t)$ depends only on $\{ X_s \}_{ 0 \leq s \leq t}$.  

For Eq.~\eqref{eqn:SDE_outcome}, the DM $\rho_t$ is regarded as a stochastic process adapted to the Brownian motion $W_t$.

\item $X(t, \{ \dd W_t \})$ represents a sample path adapted to a realization $\{ \dd W_t \}$. Key mathematical tools to analyze the asymptotic behavior are the FP equation and the Lyapunov analysis.
We mainly FP equation in the main text; the Lyapunov analysis for $B=0$ case will be discussed in Appendix \ref{app:Lyapunov}.

\item 
(Theorem) It\^{o} isometry: given $\Delta(t)$ adapted to the Brownian motion $W_t$, then $I(t) \equiv \int_0^t \Delta(s) \dd W_s$ is itself a stochastic process  and its variance is given by $\mathbb{E} \bigg[ \left(  \int_0^t \Delta(s) \dd W_s \right)^2 \bigg] = \mathbb{E} \bigg[  \int_0^t \Delta^2(s) \dd s  \bigg] $.

\ei   

%\subsection{Fokker-Planck (Chapman–Kolmogorov) equation}

\subsection{Fokker-Planck equation from SDE} 

From a general multi-variable SDE 
\beq 
\begin{aligned}
\dd x_i &= b_i (\mathbf{x}) \dd t +  \sum_{\alpha=1}^m \sigma_{i \alpha} ( \mathbf{x}) \ \dd W_\alpha (t) 
\end{aligned} 
\label{eqn:SDE_example}
\eeq 
($i$=1-$n$, $\alpha$=1-$m$) with the drift $b_i$ and white noise $\dd W_\alpha(t)$ of amplitude $\sigma_{i \alpha} $, the corresponding FP equation is 
\begin{subequations}
\begin{align}
\partial_t P (\mathbf{x} , t|\mathbf{x} _0 , t_0) &= - \frac{\partial}{\partial x_i } \cdot  [b_i (\mathbf{x} ) \ P (\mathbf{x} , t|\mathbf{x} _0, t_0 )] + \frac{\partial^2}{\partial x_i \partial x_j}   {D}_{ij} (\mathbf{x} ) P (\mathbf{x} , t|\mathbf{x} _0 , t_0)   \\
%%%%%%%%%%%
\Leftrightarrow 
\frac{\partial}{\partial t} P &= + L_\text{FP} P \text{ with }
L_\text{FP} ( \mathbf{x}  ) = -\frac{\partial}{\partial x_i }  b_i ( \mathbf{x}  ) +
\frac{\partial^2}{\partial x_i \partial x_j}   {D}_{ij} ( \mathbf{x})
\label{eqn:FP_operator_form}
\end{align}
\label{eqn:FP_operator_form_general}
\end{subequations}
In Eqs.~\eqref{eqn:FP_operator_form_general} the repeated indices are summed over, and $D_{ij} (\mathbf{x} ) = \frac{1}{2} \sum_{\alpha=1 }^m \sigma_{i \alpha} (\mathbf{x} )  \sigma_{j \alpha} (\mathbf{x} ) $ is the referred to as the diffusion matrix or diffusion coefficient(s). For SDE analyzed in this work, $n=1$ or 2 and $m=1$.

FP equation is the deterministic equation for the probability $P_{X_t}(x) = p(x,t)$. The time-independent solution of FP equation corresponds to the {\em stationary} distribution. For our applications the following features are highlighted.

\bi 

\item The stationary distribution exists when $b_i (\mathbf{x} )$ and $\sigma_{i \alpha} ( \mathbf{x})$ satisfy some continuity conditions \cite{Book_SP_Khasminskii}. For cases studied in this work we construct it numerically. 

\item If $D_{ij}$ is positive definite, any distributions asymptotically converge to the unique stationary distribution. The stationary solution of FP equation thus reveals much information about the asymptotic behavior.  

For FP equations considered in this work [Eq.~\eqref{eqn:FP_theta} and Eq.~\eqref{eqn:FP_theta_theta}], $D_{ij}$ are only non-negative definite, but it turns out that the same statement still holds once we impose continuity and periodic conditions. In Appendix \ref{app:stationary}, we outline the proof given in Ref.~\cite{book_FP_Risken} and show how additional constraints are used for non-negative definite $D_{ij}$ in our problems.

%\item Once the stationary distribution is reached, the ergodic theorem applies. 

\ei

\section{Asymptotic purity}  \label{sec:asym_purity}

%The mathematical tools summarized in the previous section are now used to analyze the numerical observations. 

The purity is characterized by $|\pmb{\rho}|^2 = \rho_x^2 + \rho_y^2 + \rho_z^2 \equiv \rho_y^2 + r^2$ [notation in Section \ref{sec:decoupling}], and the goal is to show that $1-|\pmb{\rho}|^2$ converges to zero asymptotically for all realizations.

Let us first consider the $\rho_y$ component. The SDE for $\rho_y$ is $\dd \rho_y = - \frac{\rho_y}{2} \dd t -  \rho_z \rho_y \dd W_t$; its expectation is $\mathbb{E}[\rho_y(t)] =  \rho_{y,0} e^{-t/2}$ (solving $\frac{\dd}{\dd t} \mathbb{E}[\rho_y] = - \frac{\mathbb{E}[\rho_y]}{2}$). An explicit expression of $\rho_y(t)$ can be obtained by considering the SDE of $\ln \rho_y$ via $\dd \ln \rho_y = \frac{\dd \rho_y}{\rho_y} - \frac{1}{2} \frac{(\dd \rho_y)^2}{\rho_y^2} $ :
\beq
\begin{aligned}
\dd \ln \rho_y &= -\frac{1}{2} (1+ \rho_z^2) \dd t -  \rho_z \dd W_t \\
%%%%%%
\Rightarrow \
\rho_y (t) &= \rho_{y,0} e^{-\frac{t}{2} - \frac{1}{2} \int_0^t  \rho_z^2  \dd s-  \int_0^t \rho_z \dd W_s}.
\end{aligned}
\label{eqn:rho_y2}
\eeq
Eq.~\eqref{eqn:rho_y2} indicates $\frac{ \rho_y (t) }{ \rho_{y,0} }$ is a positive random variable and without loss of generality we take $\rho_{y,0}>0$. Markov inequality $P_{\rho_y(t)} (y>y_0) \leq \frac{ \mathbb{E}[\rho_y(t)] }{y_0} = \frac{ \rho_{y,0} e^{-t/2}  }{y_0}$ grants the following $\epsilon-\delta$ statement:
\beq
\begin{aligned}
\text{$\forall$ $y_0 > 0$, $\exists$ a sufficiently long time $t >  \ln \delta^{-1} + \ln y_0^{-1} + \ln \rho_{y,0}$ such that $P_{\rho_y(t)} (y>y_0) < \delta$.}
\end{aligned}
\eeq
We thus claim $\rho_y(t)$ converges to zero in probability asymptotically.

The SDE for $r^2 = \rho_x^2 + \rho_z^2$ can be derived from Eq.~\eqref{eqn:dr_general}:
\beq 
\begin{aligned}
\dd r^2  &= 2 r\ \dd r +  (\dd r)^2  \\
&=  (1-\rho_z^2)  (1-r^2) \  \dd t  + 2  \rho_z (1-r^2)  \ \dd W_t.
\end{aligned}
\eeq 
It turns out the analysis is easier by introducing $\epsilon_t = 1 - r_t^2$ whose corresponding SDE is
\beq 
\dd \epsilon = - (1-\rho_z^2) \epsilon \  \dd t  - 2  \rho_z \epsilon \ \dd W_t. 
\label{eqn:d_epsilon'}
\eeq 
By construction $1 \geq \epsilon_t \geq 0$ for any physical states, and the state is pure when $\epsilon =0$.
To analyze the asymptotic behavior of $\epsilon_t$, we define $y_t = \ln \epsilon_t $ %and use $\dd y_t = \frac{\dd \epsilon_t}{\epsilon_t} - \frac{1}{2} \frac{(\dd \epsilon_t)^2}{\epsilon_t^2} $
to get
\beq 
\begin{aligned}
& \dd y_t = - (1 + \rho_z^2) \dd t - 2 \rho_z \dd W_t \\
%%%%%%
\Rightarrow \ & y_t = \ln \big[ \frac{\epsilon_t}{\epsilon_0} \big]
= - t -  \int_0^t \dd s \, \rho_z^2 - 2 \int_0^t \rho_z \dd W_s
< - t + \int_0^t (-2 \rho_z) \dd W_s \\
%%%%%%%%%%%%
\Rightarrow \ & \frac{ \epsilon_t }{ \epsilon_0 } =  \ e^{-t} e^{  -  \int_0^t \dd s \rho_z^2 + \int_0^t (-2 \rho_z) \dd W_s }
\leq  \ e^{-t} e^{ + \int_0^t (-2 \rho_z) \dd W_s } \equiv \bar{\epsilon}_t.
\end{aligned}
\label{eqn:purestate_explicit_02}
\eeq 
In the second expression, $\rho_z^2>0$ so that $ \int_0^t \dd s\, \rho_z^2  \geq 0$; the initial value $\epsilon_0$ is positive for any physical state. Because $0< \frac{\epsilon_t}{\epsilon_0} \leq \bar{\epsilon}_t$ at all $t$ for any realization $\{ \dd W_t \}$, $\bar{\epsilon}_t \rightarrow 0 $ implies $\frac{\epsilon_t}{\epsilon_0} \rightarrow 0$. We thus analyze $\bar{\epsilon}_t$ by computing the upper bound of $P_{\bar{\epsilon}_t }( \bar{\epsilon} > \varepsilon )$.
To proceed we define a stochastic process $\tilde{Y}_t$: %, $Y_t = e^{  \int_0^t (-2 \rho_z) \dd W_t } = e^{ \tilde{Y}_t}$, and write
\begin{subequations} 
\begin{align} 
\tilde{Y}_t &= \int_0^t (-2 \rho_z) \dd W_s,   \label{eqn:YY} \\
%%%%%%%%%%%%
\bar{\epsilon}_t &=  e^{-t} e^{ \bar{Y}_t } \equiv f^{-1}_t( \bar{Y}_t ).  \label{eqn:YY0}
\end{align}
\end{subequations} 
The variance of $\tilde{Y}_t$ can be bounded using It\^{o} isometry and $\rho_z^2<1$:
\beq 
\mathbb{E}[ ( \tilde{Y}_t )^2 ] = \int_0^t (-2\rho_z)^2 \ \dd s < 4 t.
\label{eqn:bound_Ytilde}
\eeq
To use Eq.~\eqref{eqn:bound_Ytilde} to upper bound $P_{ \bar{\epsilon}_t  } ( \bar{\epsilon} > \varepsilon )$, we consider the same distribution with $\bar{Y}_t$ as the variable $P_{ \bar{Y}_t  } ( \bar{y} )$ where  $\bar{Y}_t= f_t( \bar{\epsilon}_t ) = t + \ln \bar{\epsilon}_t$ from Eq.~\eqref{eqn:YY0}. Using Eq.~\eqref{eqn:P_variable_change}, one has
\beq
\begin{aligned}
 P_{ \bar{\epsilon}_t  } ( \bar{\epsilon} > \varepsilon )
 %&= P_{ \ln Y_t  } ( \ln y > \ln \varepsilon  + t )
 &= P_{ \tilde{ Y}_t  } ( \bar{y} > f_t(\varepsilon) )
 = P_{ \tilde{ Y}_t  } ( \bar{y} > \ln \varepsilon  + t ) \\
%%%%%%%%%%%%%
&\leq  \frac{ \mathbb{E} \big[ (\tilde{Y}_t )^2 \big]  }{ \big( \ln \varepsilon  + t \big)^2 }
\leq  \frac{ 4 t }{ \big( \ln \varepsilon  + t \big)^2 } \underset{t\rightarrow \infty}{\longrightarrow} 0.
\end{aligned}
\eeq
%To use Eq.~\eqref{eqn:bound_Ytilde} to upper bound $P_{ \bar{\epsilon}_t  } ( \bar{\epsilon} > \varepsilon )$, we note that $\bar{\epsilon} = e^{-t} y  > \varepsilon$ implies $\ln y > \ln \varepsilon  + t $ and
%\beq
%\begin{aligned}
% P_{ \bar{\epsilon}_t  } ( \bar{\epsilon} > \varepsilon )
% &= P_{ \ln Y_t  } ( \ln y > \ln \varepsilon  + t )
% = P_{ \tilde{ Y}_t  } ( y > \ln \varepsilon  + t )  \\
%%%%%%%%%%%%%
%&\leq  \frac{ \mathbb{E} \big[ (\tilde{Y}_t )^2 \big]  }{ \big( \ln \varepsilon  + t \big)^2 }  \leq  \frac{ 4 t }{ \big( \ln \varepsilon  + t \big)^2 } \underset{t\rightarrow \infty}{\longrightarrow} 0.
%\end{aligned}
%\eeq
The Chebyshev inequality is used in the second line. The time $t$ needed for $P_{ \bar{\epsilon}_t  } ( \bar{\epsilon} > \varepsilon ) < \delta$ with any small $\delta>0$ is obtained via 
\beq 
\begin{aligned}
& \frac{ 4t }{ \big( \ln \varepsilon  + t \big)^2 } < \delta  
\Rightarrow \frac{ \big( \ln \varepsilon  + t \big)^2 }{ 4 t } 
 = \frac{t}{4} + \frac{1}{2} \ln \varepsilon + \frac{(\ln \varepsilon)^2}{4 t} > \frac{1}{\delta}. 
\end{aligned}
\eeq 
Requiring $\frac{t}{4} + \frac{1}{2} \ln \varepsilon > \frac{1}{\delta}$ we get $t  > \frac{4}{\delta} + 2 \ln \frac{1}{\varepsilon}$.  
In terms of the $\delta$-$\varepsilon$ description: 
\beq 
\begin{aligned}
\text{$\forall$ $\varepsilon > 0$, $\exists$ a sufficiently long time $t > \frac{4}{\delta} + 2 \ln \frac{1}{\varepsilon}$ such that $P_{ \bar{\epsilon}_t  } ( \bar{\epsilon} > \varepsilon )  < \delta$.} 
\end{aligned}
\eeq 
We thus claim $\bar{\epsilon}_t$ (thus $\epsilon_t = 1 - r_t^2$)  converges to zero in probability asymptotically.

%The similar analysis applies to $\rho_y$ component.

%The similar analysis applies to $\rho_y$ component. The SDE for $\rho_y$ is $\dd \rho_y = - \frac{\rho_y}{2} \dd t -  \rho_z \rho_y \dd W_t$. The same procedure leads to
%\beq
%\begin{aligned}
%\dd \ln \rho_y &= -\frac{1}{2} (1+ \rho_z^2) \dd t -  \rho_z \dd W_t \\
%%%%%%
%\Rightarrow \
%\rho_y (t) &= \rho_{y,0} e^{-\frac{t}{2} - \frac{1}{2} \int_0^t \dd s \rho_z^2 -  \int_0^t \rho_z \dd W_s}  \\
%%%
%& \leq \rho_{y,0} e^{-\frac{t}{2}  + \int_0^t (-\rho_z) \dd W_s}.
%\end{aligned}
%\label{eqn:rho_y} \eeq
%with positive $\rho_{y,0} $. Eq.~\eqref{eqn:rho_y} and Eq.~\eqref{eqn:purestate_explicit_02}  have the same structure. Following the same steps we conclude that $\rho_y (t) \rightarrow 0$ asymptotically in probability.

As both $\epsilon_t = 1 -\rho_x^2 - \rho_z^2 $ and $\rho_y$ both approach zero asymptotically in probability, we thus prove that the single-qubit quantum trajectory becomes pure asymptotically no matter what the initial condition is. Some general remarks will be provided in Section \ref{sec:general_remarks}.

%\subsection{initial-state dependence}  

%%%%%%%%%%%%%%%%%%%%%%%%%%%%%%%
\section{Initial-state dependence} 

We now discuss the observation that all quantum trajectories evolve into the same realization specific pure state asymptotically. As Section \ref{sec:asym_purity} establishes the asymptotic purity, the following analysis will be confined to the pure initial states. % and asymptotically converge the same one.

\subsection{Process of two initial conditions}  \label{sec:FP_1_2_variables}

Consider two pure states described by the angular variables $\theta$ and $\bar{\theta}$  that satisfy \eqref{eqn:dtheta_pure} with same realization. The SDE for  $(\theta, \bar{\theta})$ is 
\begin{equation}
\begin{aligned} 
 %%%%%%%
 \dd  \begin{bmatrix} \theta \\ \bar{\theta} \end{bmatrix} &=  \begin{bmatrix} B + \frac{1}{4} \sin (2 \theta) \\ B + \frac{1}{4} \sin (2 \bar{\theta})    \end{bmatrix}  \dd t + 
 \begin{bmatrix}  \cos \theta \\ \cos \bar{\theta} \end{bmatrix}  \dd W_t .
 \label{eqn:1D_couple_origin_3}
\end{aligned} 
\end{equation}   
The FP equation for $\theta$ alone is
\beq 
\frac{\partial }{\partial t}P^{(1)}(\theta, t) = 
- \frac{\partial}{\partial \theta} \big[P^{(1)} (B+\frac{1}{4} \sin(2 \theta) ) \big] + \frac{1}{2} \frac{\partial^2}{\partial \theta^2} \big[P^{(1)} \cos^2 \theta \big] .
\label{eqn:FP_theta}
\eeq  
The FP equation for $(\theta, \bar{\theta})$ is given by 
the FP equation for joint probability $P^{(2)}(\theta, \bar{\theta}, t)$ is 
\begin{equation}
\begin{aligned} 
\frac{\partial }{\partial t}P^{(2)}(\theta, \bar{\theta}, t) &=
 - \frac{\partial}{\partial \theta} \big[(B+ \frac{1}{4} \sin(2 \theta) ) P^{(2)}  \big]
 - \frac{\partial}{\partial \bar{\theta} } \big[(B+ \frac{1}{4} \sin(2 \bar{\theta} ) ) P^{(2)}  \big] \\ 
 %%%%%%%%%%%%%%%%
&+ \frac{1}{2} \begin{pmatrix*}[l]
 + \frac{\partial^2}{\partial \theta^2} \big[ \cos^2( \theta) \ P^{(2)}\big] \\
+ 2 \frac{\partial^2}{\partial \theta \partial \bar{\theta}}  
 \big[ \cos(\theta) \cos(\bar{\theta})  P^{(2)}\big]
\\ 
+ \frac{\partial^2}{\partial \bar{\theta}^2} \big[ \cos^2( \bar{\theta} ) \ P^{(2)}\big]       
\end{pmatrix*} 
\equiv L_\text{FP}(\theta, \bar{\theta} ) P^{(2)}.
\end{aligned} 
\label{eqn:FP_theta_theta}
\end{equation}  

Let us assume the existence of the stationary distribution $P^{(1)}_\text{sta}(\theta)$ that satisfies Eq.~\eqref{eqn:FP_theta}, i.e., 
\beq 
- \frac{\partial}{\partial \theta} \big[P_\text{sta}^{(1)} (B+\frac{1}{4} \sin(2 \theta) ) \big] + \frac{1}{2} \frac{\partial^2}{\partial \theta^2} \big[P_\text{sta}^{(1)} \cos^2 \theta \big] = 0, 
\label{eqn:FP_theta_stationary}
\eeq 
and discuss its consequences.  First, if $P^{(1)}_\text{sta}(\theta)$ is non-zero over the entire domain, then it can be shown that any  $P^{(1)} (\theta, t)$ eventually converges to the stationary $P^{(1)}_\text{sta}(\theta)$ [see Appendix \ref{app:stationary}]. Therefore within one realization $\theta_t$ is moving around the entire $(-\pi, \pi]$ with the frequency inside $[\theta, \theta+\dd \theta]$ proportional to $P^{(1)}_\text{sta}(\theta) \dd \theta$ (ergodicity).  
If $P^{(1)}_\text{sta}(\theta)$ is zero over some finite domain, then within one realization $\theta_t$ has to be ''trapped`` in one of disjoint domains because the stochastic process $\theta_t$ is {\em continuous} and $\theta_t$ cannot  move from $a-y$ to $a+y$ without passing $a$;  $B=0$ corresponds to this case.
$P^{(1)}_\text{sta}(\theta)$ will be explicitly constructed  in Section \ref{sec:P_stat_FT_CF}.

Given $P^{(1)}_\text{sta}$, it can be shown that $P^{(2)}_\text{sta}(\theta, \bar{\theta}) = P^{(1)}_\text{sta}(\theta) \delta(\theta - \bar{\theta}) = P^{(1)}_\text{sta}(\bar{\theta}) \delta(\theta - \bar{\theta})$ satisfies $L_\text{FP}(\theta, \bar{\theta} ) P^{(2)}_\text{sta}=0$ (see Appendix \ref{app:2v_sta} for a proof in the weak form). 
Notice that $\theta$ and $\bar{\theta}$ are {\em not} independent because of the same realization; in terms of joint probability, $P^{(2)}_\text{sta}(\theta, \bar{\theta}) \neq  P^{(1)}_\text{sta}({\theta}) P^{(1)}_\text{sta}(\bar{\theta})$.
Marginalization of $P^{(2)}_\text{sta}(\theta, \bar{\theta})$ recovers the stationary distribution of $\theta$: $  P^{(1)}_\text{sta}(\theta) = \int \dd \theta \ P^{(2)}_\text{sta}(\theta, \bar{\theta})$.
As any $P^{(2)}(\theta, \bar{\theta}, t)$ converges to $P^{(2)}_\text{sta}(\theta, \bar{\theta}) = P^{(1)}_\text{sta}(\theta) \delta(\theta - \bar{\theta})$ at large times (see Appendix \ref{app:stationary}), we thus conclude that any two quantum trajectories under the same realization $\{ \dd W_t\}$ coincide asymptotically. 
%[so far we only have a stationary distribution but cannot say any initial distribution will converge to that distribution!]

\subsection{$P^{(1)}_\text{sta}(\theta)$ construction using  continued fraction} \label{sec:P_stat_FT_CF}

\begin{figure}[ht]
\centering
\includegraphics[width=0.8\textwidth]{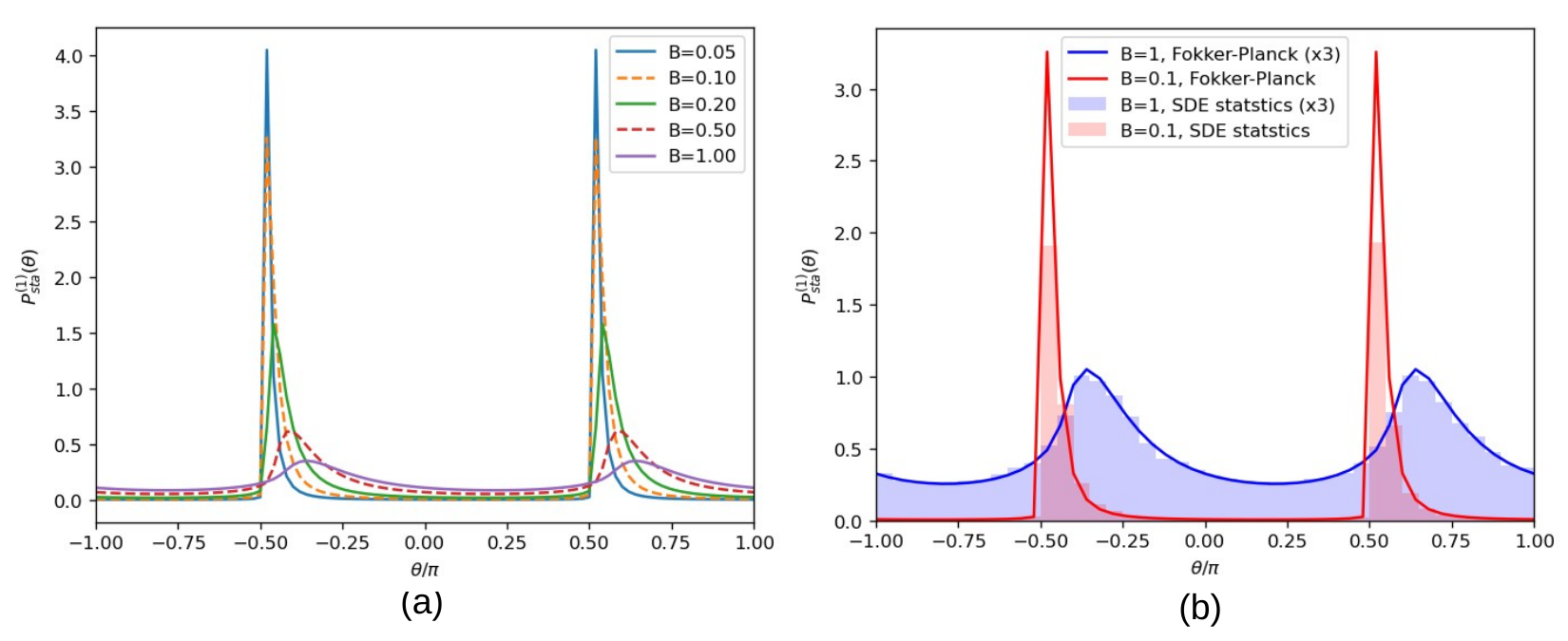}
\caption{Stationary probability distribution for a single qubit. (a) Results from time-independent Fokker-Planck equation using continued fraction method. (b) Demonstrating the ergodic property using $B=0.1$ and $B=1$. The histograms (bars), which are very close to $P^{(1)}_\text{sta}(\theta; B)$, are based on SDE simulations of the same $B$. The results of $B=1$ are multiplied by 3 for better visualization.
}
\label{fig:StationaryP_1Q}
\end{figure}

The stationary distribution is now explicitly constructed. 
Due to the periodic boundary condition, the solution of Eq.~\eqref{eqn:FP_theta_stationary} can be expressed using Fourier series $P^{(1)}_\text{sta}(\theta) = \sum_{n=-\infty}^\infty c_n e^{-i n \theta}$. Substituting 
\beq 
\begin{aligned}
 \sin(2 \theta) P &= \sum_m  \frac{ c_{m+2} - c_{m-2} }{2i} e^{-i m \theta},  \\
%%%%%%%%%%%%%
(\cos(2 \theta) + 1 )P &=  \sum_m  \bigg[ \frac{ c_{m+2} + c_{m-2} }{2} + c_m \bigg] e^{-i m \theta}  
\end{aligned}
\nonumber 
\eeq 
into Eq.~\eqref{eqn:FP_theta} we get
\beq  
\begin{aligned}
%&  (i B - \frac{m}{4} ) a_m - \frac{m-1}{8}  a_{m+2}  - \frac{m+1}{8} = 0.  \\
%%%%%%%%%%%%
%\Leftrightarrow \ &  
\underbrace{ \big( 1 - \frac{i 4 B}{m} \big) }_{ \equiv {Q}_m } c_m + \underbrace{  \frac{1}{2} (1 - \frac{1}{m}) }_{ \equiv {Q}^+_m }  c_{m+2} + \underbrace{ \frac{1}{2} (1 + \frac{1}{m})  }_{ \equiv {Q}^-_m } c_{m-2}= 0.
\end{aligned}
\label{eqn:iteration}
\eeq  

Coefficients ${Q}_m =  1 - i \frac{4 B}{m}$, ${Q}^+_m = \frac{1}{2}(1 - \frac{1}{m})$, and ${Q}^-_m = \frac{1}{2}(1 + \frac{1}{m})$ are defined. The recursion relation Eq.~\eqref{eqn:iteration} holds for $m \neq 0$, and the normalization condition requires $c_0 = \frac{1}{2 \pi}$.
Following Chapter 9 of Ref.~\cite{book_FP_Risken}, a stable numerical scheme is obtained by introducing the quotient $S_m = \frac{c_{m+2}}{ c_m }$ whose  recursion relation $\frac{Q^-_{m+2} }{ S_{m} } + Q_{m+2} + Q^+_{m+2} S_{m+2} = 0$ (taking $m \rightarrow m+2$ in Eq.~\eqref{eqn:iteration} and then divided by $c_{m+2}$) leads to the continued fraction
\beq 
\begin{aligned}
% & , \text{ or }   \frac{Q^-_{m+2} }{ S_{m} } + Q_{m+2} + Q^+_{m+2} S_{m+2} = 0.  \\
%%%%%%%%%%%%%%%%%
%\Rightarrow &  
S_m = \frac{ - Q^-_{m+2}  }{ Q_{m+2} + Q^+_{m+2} S_{m+2}  } 
= \cfrac{  - Q^-_{m+2}  }{ Q_{m+2} - \cfrac{ Q^+_{m+2} Q^-_{m+4}  }{ Q_{m+4} 
  - \cfrac{Q^+_{m+4} Q^-_{m+6} }{Q_{m+6} - \cdots}   }  } .
\end{aligned}
\label{eqn:a_n_FP_SingleQubit}
\eeq 
The corresponding list for partial numerators, denoted as $\mathbf{a}_m $, and that for partial denominators, denoted as $\mathbf{b}_m $ are respectively [see Appendix \ref{app:CF} for terminology]
\beq 
\begin{aligned}
\mathbf{{a} }_m &= [ -Q^-_{m+2}, -Q^+_{m+2}Q^-_{m+4}, -Q^+_{m+4}Q^-_{m+6}, \cdots  ]   \\
&= \bigg[ -\frac{1}{2} \frac{m+3}{m+2}, - \frac{ (m+1)(m+5) }{4(m+2)(m+4)}, - \frac{ (m+3)(m+7) }{4(m+4)(m+6)}, - \frac{ (m+5)(m+9) }{4(m+6)(m+8)},  \cdots \bigg], \\
%%%%%%%%%%%%%%%%%%%
\mathbf{ {b} }_m  &= [ Q_{m+2}, Q_{m+4}, Q_{m+6}, \cdots  ] 
= \bigg[ 1 - \frac{i 4 B}{m+2}, 1 - \frac{i 4 B}{m+4}, 1 - \frac{i 4 B}{m+6}, \cdots  \bigg].
\end{aligned}
\eeq 
For $B=0$, we can prove that the stationary solution is the sum of two $\delta$-function, i.e., 
\beq 
P^{(1)}_\text{sta} (\theta; B=0) = \frac{1}{2} \big[ \delta( \theta - \frac{\pi}{2}) +  \delta( \theta + \frac{\pi}{2}) \big]. 
\label{eqn:Pstat_B0}
\eeq 
This is done by showing that $S_{2m} = -1$ so that $c_{2m} = \frac{ (-1)^m }{2 \pi}$. Details are provided in Appendix \ref{app:CF}. 

Fig.~\ref{fig:StationaryP_1Q}(a) shows the stationary distribution for a few $B$ values. To numerically confirm the ergodic property for non-zero $B$'s,
we collect $\theta_t$'s from an arbitrary initial condition and one realization $\{ \dd W_t \}$ and compare its histogram of $\theta_t$ with the corresponding $P^{(1)}_\text{sta}(\theta; B)$. Ergodicity implies
\beq 
P^{(1)}_\text{sta}(\theta) \dd \theta 
\propto \text{number of $\theta_t$'s in $[\theta, \theta+\dd \theta]$} .
\label{eqn:ergodic}
\eeq 
Eq.~\eqref{eqn:ergodic} is numerically tested, and results for $B=0.1$ and $B=1$ are shown in Fig.~\ref{fig:StationaryP_1Q}(b). In the calculations shown here, we keep 500 positive Fourier components and use 100th approximant of continued fraction [Eq.~\eqref{eqn:approximant}]; we have checked that using more terms makes negligible differences.  
 
\subsection{Expectation and probability current}  % (and transition rate??)

\begin{figure}[ht]
\centering
\includegraphics[width=0.8\textwidth]{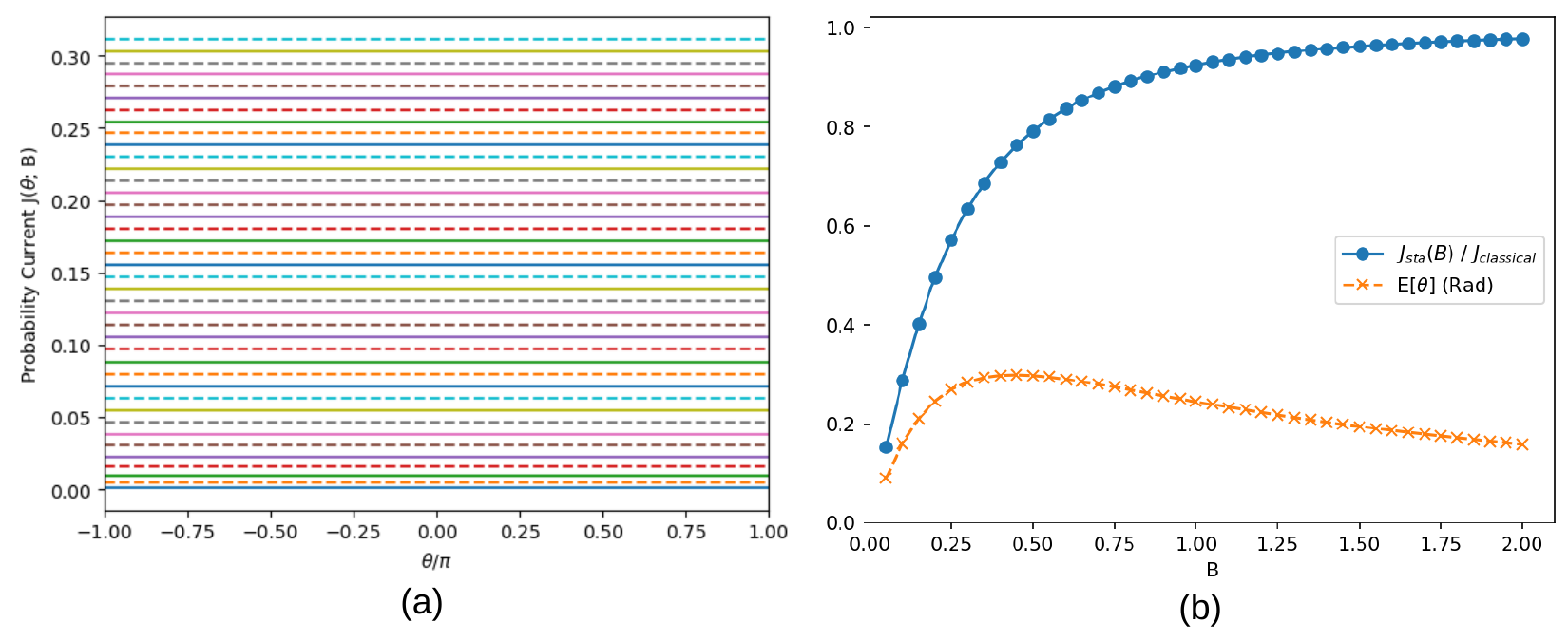}
\caption{Probability current. (a) $J_\text{sta}(\theta; B)$ for $B=0.05$ to 2. The flatness is the consequence of stationary distribution. Curves from bottom to top correspond to $B=0.05 + n\cdot 0.05$ with $n=0$ to $n=39$.  (b) solid curve: $J_\text{sta}(B)/J_\text{classical}$; dashed curve: $\mathbb{E}_\text{sta}[\theta]$ for the same range of $B$. 
}
\label{fig:StationaryP_Cur_1Q}
\end{figure}

%The probability current is defined by 
Equation of continuity $\partial_t P = - \partial_\theta J$ defines a probability current $J = P^{(1)}(\theta) (B + \frac{1}{4} \sin(2 \theta)) - \frac{1}{4} \frac{\partial}{\partial \theta} 
\big[  P^{(1)} (\cos(2 \theta)+1) \big]$. With the stationary distribution one gets 
\beq 
\begin{aligned}
J_\text{sta}(\theta; B) &= P^{(1)}_\text{sta}(\theta) (B + \frac{1}{4} \sin(2 \theta)) - \frac{1}{4} \frac{\partial}{\partial \theta} 
\big[  P^{(1)}_\text{sta}(\theta) (\cos(2 \theta)+1) \big] \\
%%%%%%%%%%%%%5
& \rightarrow 
\sum_m \big[ B + \frac{1}{4} \sin(2 \theta) \big] c_m e^{-i m \theta} 
+\frac{1}{4} \sum_m \bigg[ i m [ \cos (2 \theta)+1 ] + 2 \sin(2 \theta) \bigg] c_m e^{-i m \theta}  \\
%%%%
&= \sum_m \bigg[ \frac{ i m }{4} [ \cos (2 \theta)+1 ] + \frac{3}{4} \sin(2 \theta) + B \bigg] c_m e^{-i m \theta}  \\
%%%%%%%%%%
& \underbrace{ \longrightarrow }_{\theta \text{-ind.}} 
B c_0 + \frac{\text{Im}[c_2] }{4} =  c_0 B\big( 1 + \frac{\text{Im}[S_0] }{4B} \big)
\end{aligned}
\label{eqn:Prob_cur_stationary}
\eeq 
For stationary distribution $J_\text{sta}(\theta; B)$ is $\theta$-independent [see Fig.~\ref{fig:StationaryP_Cur_1Q}(a)] and the evaluation of $J_\text{sta}(B)$ only involves $S_0$ (or $c_0 = \frac{1}{2 \pi}$ and $c_2$) as expressed in the last expression of Eq.~\eqref{eqn:Prob_cur_stationary}.
%which is numerically tested in . 
In the classical limit where $B$-field dominates, $P_\text{sta}(\theta) = \frac{1}{2 \pi}$ and $J_\text{classical} = \frac{B}{2 \pi}$. As the presence of diffusion generates the resistance to the motion of $\theta_t$, one expects $J_\text{sta}(B) \leq J_\text{classical}$.  
When $B=0$, $S_0=-1$ so  $J_\text{sta}(B=0)=0$. 
Fig.~\ref{fig:StationaryP_Cur_1Q}(b) plots $\frac{J_\text{sta}(B)}{J_\text{classical}}$, which indeed approaches one from below upon increasing $B$.
%Selected results are given in Fig.~\ref{fig:StationaryP_Cur_1Q}.

The expectation of $\theta$ over the stationary distribution is 
\beq 
\begin{aligned}
\mathbb{E}_\text{sta}[\theta] &= \int_{-\pi}^{\pi} \dd \theta \ \theta  P^{(1)}_\text{sta}(\theta) = - 2 \pi \sum_{m=1}^\infty \frac{\text{Im}[c_{2m}]}{m}.  
\end{aligned}
\label{eqn:mean_theta}
\eeq 
When $B$ is small, $\mathbb{E}_\text{sta}[\theta]$ is small because $P^{(1)}_\text{sta}(\theta)$ has two peaks located approximately symmetrically around $\theta = \pm \pi/2$. When $B$ is large, $\mathbb{E}_\text{sta}[\theta]$ is also small because $\theta  P^{(1)}_\text{sta}(\theta)$ becomes more uniform. 
The values for $B=0.05$ to 2 are plotted in Fig.~\ref{fig:StationaryP_Cur_1Q}(b) [dashed curve]. As expected $\mathbb{E}_\text{sta}[\theta]$ exhibits a maximum around $B \approx 0.4$ and approaches zero for small and large $B$. % This is consistent with

\subsection{Experimental implications} \label{sec:expt_implication}

\begin{figure}[ht]
\centering
\includegraphics[width=0.8\textwidth]{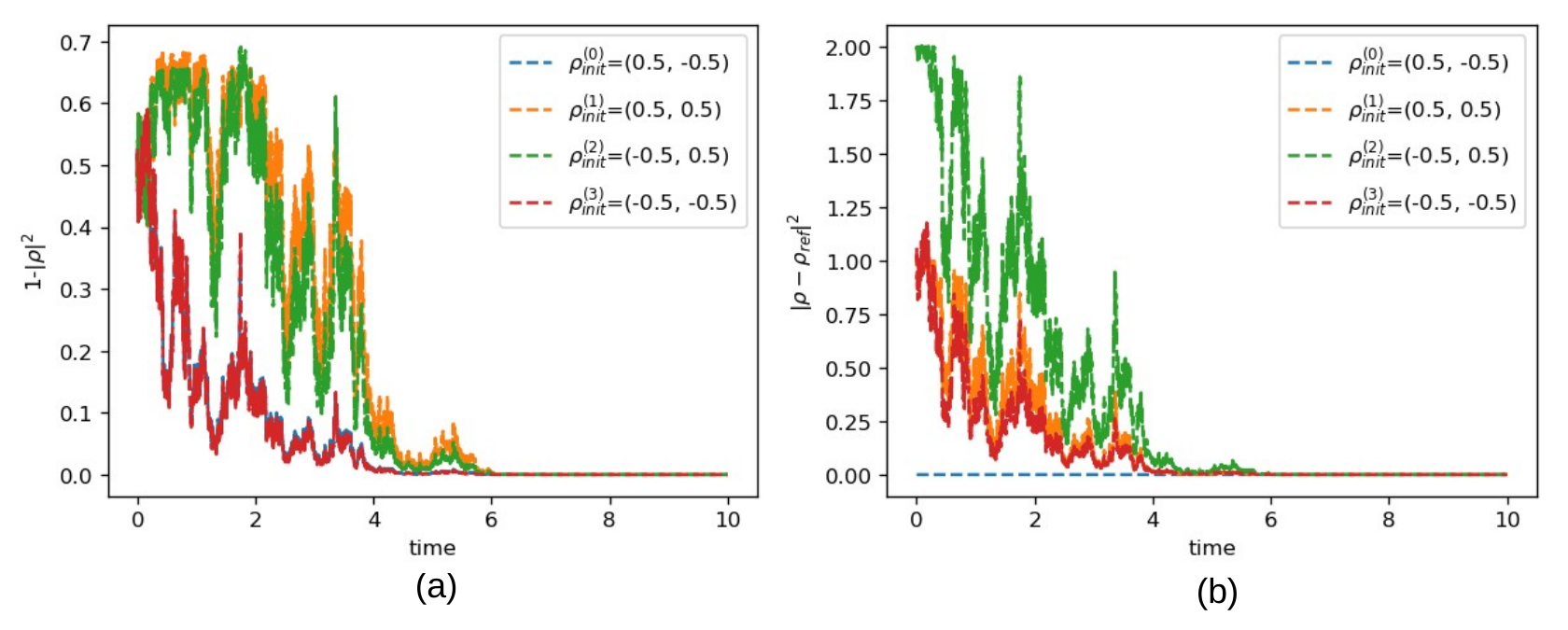}
\caption{ 
Constructing the measurement outcomes $\{ \dd Y_t \}_{\pmb{\rho}^{(0)}_\text{init} }$ from the initial state $\pmb{\rho}^{(0)}_\text{init}  = (0.5, -0.5)$ and evolving the system using $\{ \dd Y_t \}_\text{ref}$ from different initial states. $B=0.1$ is used in the simulation. (a) The states become pure asymptotically. (b) All initial states converge to the same the same asymptotic state. $|\pmb{\rho}_1 - \pmb{\rho}_2|$ is the Euclidean distance between two Cartesian vectors.
}
\label{fig:1Qubit_dYt_diff_init}
\end{figure} 

For a given realization $\{ \dd W_t \}$, let us consider the following three processes for a single qubit [see Eq.~\eqref{eqn:SME_1Qubit}]: 
\beq
\begin{aligned}
&  \rho^{(0)}(t + \dd t) = \rho^{(0)}(t) + f[ \rho^{(0)}(t)] \dd t + g[\rho^{(0)}(t)] \dd W_t, \ \  \dd W_t = \dd Y^{(0)}_t - \rho^{(0)}_z(t) \dd t, \\
  %%%%%%%%
  & \rho^{(1)}(t + \dd t)= \rho^{(1)}(t) + f[ \rho^{(1)}(t)] \dd t + g[\rho^{(1)}(t)] \dd W_t, \ \  \dd W_t = \dd Y^{(1)}_t - \rho^{(1)}_z(t) \dd t,  \\
  %%%%%%%%%%%%%%
  & \rho^{(2)}(t + \dd t) = \rho^{(2)}(t) + f[ \rho^{(2)}(t)] \dd t + g[\rho^{(2)}(t)] \cdot \big[ \underbrace{ \dd W_t + \rho^{(0)}_z(t) \dd t }_{ = \dd Y^{(0)}_t} - \rho^{(2)}_z(t) \dd t \big] ,
 \\
 & \text{where }   \ \ \rho^{(0)}(0) \neq \rho^{(1)}(0) = \rho^{(2)}(0) .
 \end{aligned}
\label{eqn:3_processes}
\eeq 
Our analysis so far concerns the processes $\rho^{(0)}(t)$ and $\rho^{(1)}(t)$, indicating $\rho^{(1)}(t) \rightarrow \rho^{(0)}(t)$ as $t \rightarrow \infty$. The corresponding implementation is as follows: preparing two different initial states and evolving them with the {\em same} realization $\{\dd W_t\}$, and we shall find that they converge to the same pure state. However this setup cannot be realized because experimentally it is impossible (in the sense of measure-zero) to produce two identical realizations; moreover the experimental outcomes are $\dd Y_t$ not directly $\dd W_t$.

To make connection to experiments let us compare $\rho^{(0)}(t)$ with $\rho^{(2)}(t)$. $\rho^{(0)} (t)$ is the reference trajectory that is used to generate the experimental outcomes $\{ \dd Y^{(0)}_t \}$ whereas $\rho^{(2)}(t)$ is updated using  $\{ \dd Y^{(0)}_t \}$.  If we {\em assume} both $\rho^{(0)}(t)$ with $\rho^{(2)}(t)$ are pure states (we only prove $\rho^{(0)}(t)$ becomes pure, not $\rho^{(2)}(t)$), following the same procedure of Section \ref{sec:FP_1_2_variables}  we can show that $\rho^{(2)}(t)$ eventually agrees with $\rho^{(0)}(t)$  for single-qubit systems. Some details are provided in Appendix \ref{app:expt_implication}. Our simulations suggest that this phenomenon holds for mixed initial states as well.
Experimentally this implies that even one starts with the wrong initial state, continuous updating using reference experimental outcomes eventually brings the system to the reference trajectory.

%We show $\rho^{(2)}(t)$ agrees with $\rho^{(1)}(t)$ and thus $\rho^{(0)}(t)$. Define $\Delta \rho_z (t) = \rho^{(0)}_z (t)- \rho^{(2)}_z (t)$

To illustrate we construct the measurement outcomes $\{ \dd Y_t \}_{{\rho}_0}$ from a given initial state ${\rho}_0$ and the resulting reference quantum trajectory is denoted as ${\rho}_t({\rho}_0; \{ \dd Y_t \}_{{\rho}_0} )$.  
According to our simulations, any quantum trajectories updated using $\{ \dd Y_t \}_{{\rho}_0}$ indeed converge to ${\rho}_t({\rho}_0)$ eventually. In the single-qubit example shown in Fig.~\ref{fig:1Qubit_dYt_diff_init}, 
the same four initial states (i.e., $\pmb{\rho}^{(i)}_\text{init}$, $i$=0,1,2,3) are considered in Fig.~\ref{fig:1Qubit_main}. The reference quantum trajectory is  $\pmb{\rho}_t(\pmb{\rho}^{(0)}_\text{init}; \{ \dd Y_t \}_{\pmb{\rho}^{(0)}_\text{init}})$ is generated using $B=0.1$. Fig.~\ref{fig:1Qubit_dYt_diff_init}(a) shows that all $\pmb{\rho}_t(\pmb{\rho}^{(i)}_\text{init}; \{ \dd Y_t \}_{\pmb{\rho}^{(0)}_\text{init}})$'s  become pure asymptotically; Fig.~\ref{fig:1Qubit_dYt_diff_init}(a) shows that all $\pmb{\rho}_t(\pmb{\rho}^{(i)}_\text{init}; \{ \dd Y_t \}_{\pmb{\rho}^{(0)}_\text{init}})$'s become identical to the reference trajectory asymptotically. 

%\subsection{System of many qubits}   

\begin{figure}[ht]
\centering
\includegraphics[width=0.8\textwidth]{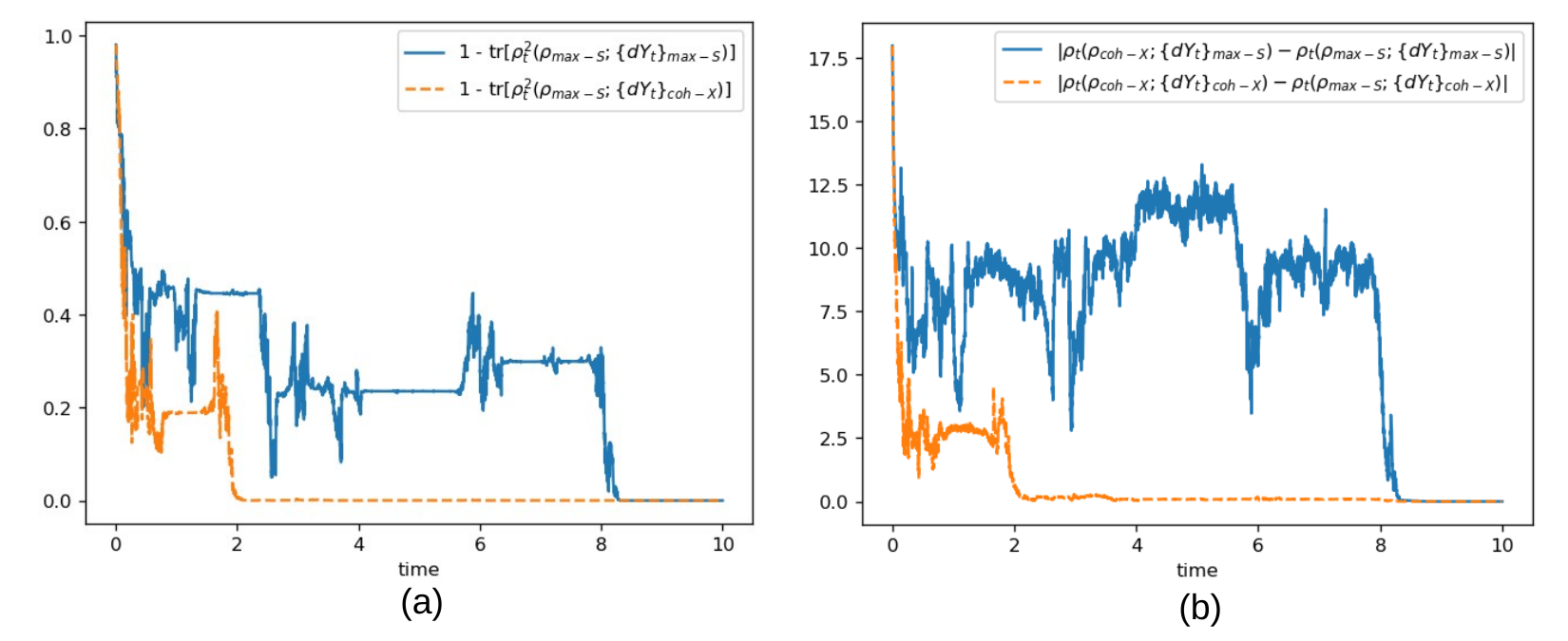}
\caption{Results of 50 qubits starting from a coherent state (maximum-eigenvalue state of $\hat{J}_x$) and from a maximum-entropy state ($\rho_\text{init} \sim \mathbb{I}$). $B=5$ is used. 
$\rho_t (\rho_0; \{ \dd Y_t \})$: the density matrix at time $t$ starting from $\rho_0$ and being updated using experimental outcomes $\{ \dd Y_t \}$.
In (a) and (b), solid curves correspond to $\{ \dd Y_t\}_\text{max-S}$; dashed curves to $\{ \dd Y_t\}_\text{coh}$.
(a) The maximum-entropy state becomes pure. 
(b) Two quantum trajectories coincide. $|{\rho}_1 - {\rho}_2| \equiv \sum_{ij} | {\rho}_{1, ij} - {\rho}_{2, ij} |$. 
}
\label{fig:50_Qubit}
\end{figure}

Fig.~\ref{fig:50_Qubit} we repeat the analysis for a system of 50 qubits with $B=5$. Two initial states are considered: the coherent state (coh) that is the eigenstate of maximum-eigenvalue of $\hat{J}_x$ and the maximum-entropy (max-S) state where the DM is proportional to the identity matrix. 
After generating two lists experimental outcomes $\{ \dd Y_t\}_\text{coh}$ and $\{ \dd Y_t\}_\text{max-S}$, we evolve the quantum trajectories from both initial states using both $\{ \dd Y_t\}_\text{coh}$ and $\{ \dd Y_t\}_\text{max-S}$. $\rho_t (\rho_0; \{ \dd Y_t \})$ represents the density matrix at time $t$ starting from $\rho_0$ and being updated using $\{ \dd Y_t \}$.
Fig.~\ref{fig:50_Qubit}(a) shows that the quantum trajectory starting from maximum-entropy state becomes pure for both $\{ \dd Y_t\}_\text{coh}$ and $\{ \dd Y_t\}_\text{max-S}$; Fig.~\ref{fig:50_Qubit}(b) shows that quantum trajectories starting from both initial states eventually coincide. We have repeated the calculations using many realizations, various initial states and for different number of qubits (up to 200), and found this phenomenon very robust. 
%In short, it appears that the quantum trajectories using the same experimental outcomes asymptotically coincide.

\subsection{General remarks and summary} \label{sec:general_remarks}

We conclude this section by comparing our results with existing works and pinpointing the differences.
It is well established that QND measurement eventually brings the system to a time-independent measurement eigenstate \cite{PhysRevA.60.2346, PhysRevLett.85.1594, book_Wiseman_Milburn}. The convergence rate is shown to be exponential \cite{Benoist-2014}, which can be understood from the decay of the off-diagonal elements of $\rho_t$ expressed in the measurement eigenstates \cite{book_Wiseman_Milburn}. There are two feedback controls that are related to the asymptotic purity. First, in the framework of QND it has been shown that the feedback control can guide the system to the target measurement eigenstate \cite{van_Handel_2005, Benoist-2014, 8618681}. Second, with the feedback control that aims to keep eigenbases of $\rho_t$ and of those of measurement {\em unbiased} \cite{PhysRevLett.96.010504}, the dynamics of $1-\text{tr}[\rho_t^2]$ becomes deterministic and the purification process can be accelerated  \cite{PhysRevA.63.062305, PhysRevA.67.030301, PhysRevLett.96.010504, Wiseman_2006,  PhysRevA.82.022307, Ruskov_2012}. The magnetometer setup considered here is generally {\em not} QND and is without feedback control, therefore the asymptotic pure state keeps on changing in time and the dynamics of $1-\text{tr}[\rho_t^2]$ remains stochastic asymptotically. Summarizing these results, it appears that the asymptotic purity is a general feature of the weak continuous measurements no matter what the unitary evolution is. The unitary evolution can be of QND type where the system Hamiltonian commutes with measurement \cite{PhysRevA.60.2346, PhysRevLett.85.1594}, conditioned on measurements to either reach a target measurement eigenstate \cite{8618681} or to accelerate the purification \cite{PhysRevLett.96.010504}, or simply generic where the system Hamiltonian does not commute with measurement (results in Section \ref{sec:asym_purity}).
As a technical comment, in the former two scenarios the measurement eigenstates are the natural bases for analysis, whereas the last scenario has no obvious preferred bases.

%Our analysis in Section \ref{sec:asym_purity} further implies that the asymptotic purity is a general feature of the weak continuous measurements; the unitary evolution commutes with measurement operator (QND) or not, is conditioned on the measurements, or a generic non-QND.

The main conclusion of this Section is that, within magnetometer setup the asymptotic pure state (without feedback control, still stochastic) depends only on the realization $\{ \dd W_t \}$, not on the initial state. This statement is proved for the single-qubit case.
Numerical simulations further strongly support that the asymptotic pure state depends only on the measurement outcomes $\{ \dd Y_t \}$, and for single qubit we provide a partial proof by assuming pure initial states. To our knowledge there is no general proof for this statement, and a somehow mathematically heavy proof can be found in Ref.~\cite{doi:10.1142/S0219025709003549} assuming nondemolition conditions. A very insightful argument built upon the asymptotic purity is provided by Jacobs in Ref.~\cite{book:Jacobs}, which may be best described using  the evolution of the un-normalized DM \cite{PhysRevA.50.5242, book_Wiseman_Milburn}. Given measurement outcomes $\{ \dd Y_t \}$, the un-normalized DM, denoted as $\tilde{\rho}_t$, is evolving according to
\beq
\tilde{\rho}_T = \bar{\Omega}( \{ \dd Y_t \}   ) \rho_0 \bar{\Omega}^\dagger( \{ \dd Y_t \}   ) \underset{  \text{pure}  }{\longrightarrow} A | \psi_T \rangle \langle \psi_T |,
\label{eqn:tilde_rho_dy}
\eeq
where $\bar{\Omega}( \{ \dd Y_t \}) = \Omega (\dd Y_{t_N} ) \Omega (\dd Y_{t_{N-1} } ) \cdots \Omega (\dd Y_{t_1} ) $ with $\Omega (\dd Y_{t} ) $ the un-normalized Kraus operator
\beq
\begin{aligned}
\Omega (\dd Y_t) &=\mathbb{I} - i B \hat{J}_y \,\dd t - \frac{1}{2} \hat{J}_z^2 \, \dd t + \hat{J}_z \dd Y_t, %, \text{ with} \\
%%%%%%
%& \overline{\dd Y_t} = 2 \text{tr}\big[ \hat{J}_z \rho \big] dt, \,\,\, \overline{dY^2_t} = dt.
 \end{aligned}
 \label{eqn:Kraus_unnormalized}
\eeq
Because the asymptotic state is pure, $\bar{\Omega}( \{ \dd Y_t \})$ has to be a projector up to a prefactor. Because $\bar{\Omega}( \{ \dd Y_t \})$ has no $\rho_0$ dependence, the projector $\bar{\Omega}( \{ \dd Y_t \}) \sim | \psi_T \rangle \langle \psi_T |$ cannot depend on $\rho_0$ which implies that the asymptotic state is independent of the initial state. %by treating infinitely many weak measurements as a {\em projection} operator to a pure state, the resulting asymptotic state has to be independent of the initial states as far as the overlap is non-zero.
Indeed in our single-qubit proof, once restricted to the pure state the analysis is straightforward.
%This may be expected but to our knowledge has not been explicitly stated.
%To our knowledge there is no explicit statement.
%As an illustration and an application
One may wonder if the insensitivity to the initial state has any consequences for the magnetometer. We believe it allows the real-time parameter estimation which will be detailed in next Section. %consider the  whose feasibility replies on . % \ref{sec:online_PE}

%\change{ }

%%%%%%%%%%
\section{ Real-time parameter estimation } \label{sec:online_PE}

\subsection{Overview}

\begin{figure}[ht]
\centering
\includegraphics[width=0.75\textwidth]{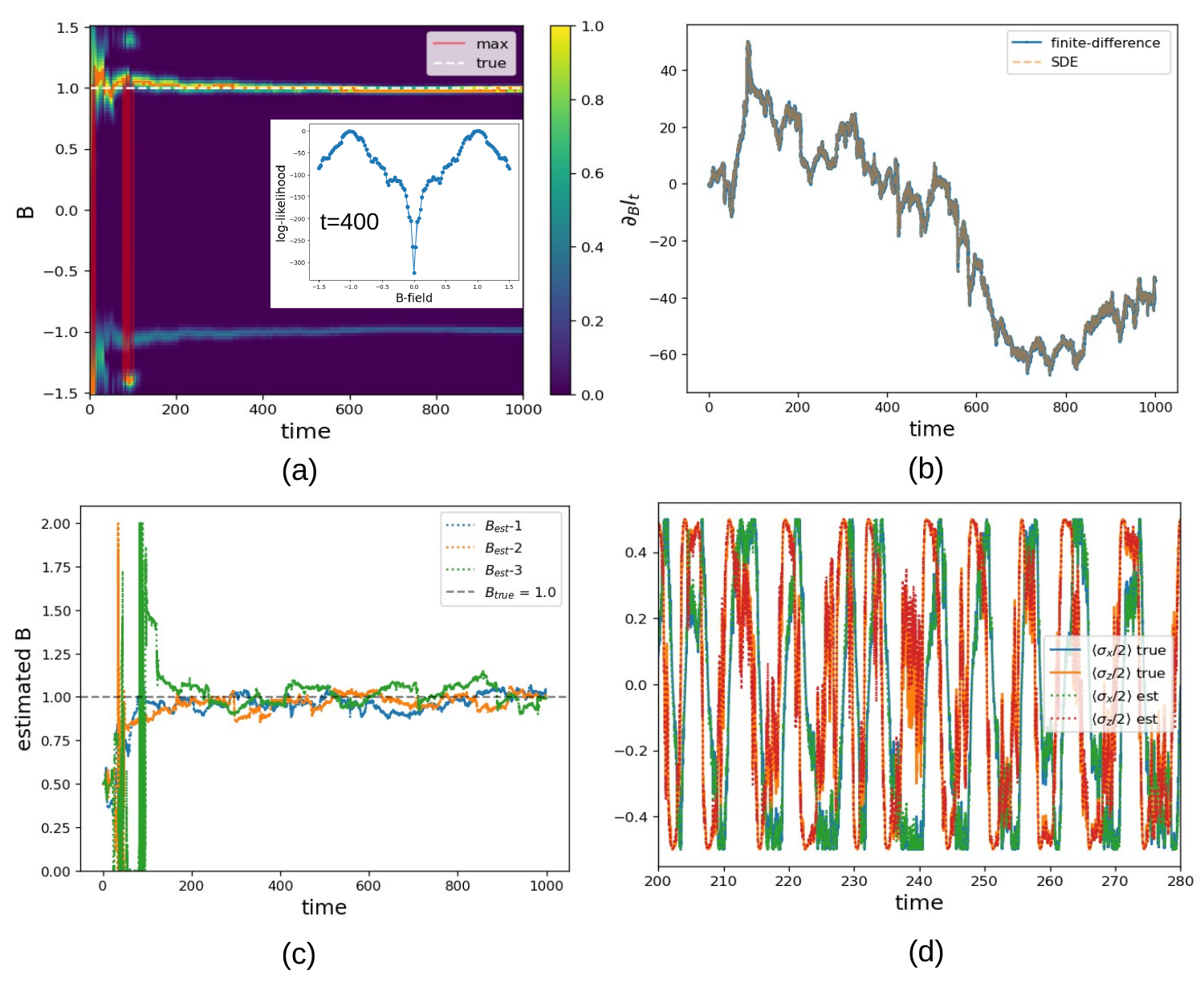}
\caption{ Real-time parameter estimation. Results shown here are obtained using $N=1$, $B_\text{true}=1$, and $\dd t=0.01$.
(a) Likelihood function $L_t(B)$. The normalization is such that $\max L_t(B)=1$. There is a local maximum at $B = -B_\text{true}$. The inset shows the log-likelihood $l_t(B)$ at $t=400$ [Eq.~\eqref{eqn:dl_t_correct}].  (b) $\partial_B l_t(B) \equiv l_t^B$ evaluated using its SDE and from the finite-difference approximation [\eqref{eqn:dl^B_dt_FD} with $\dd B = 0.02$]; they are practically identical.
(c) Estimated $B$ with three realizations, with the prior knowledge $B \in [0,2]$. $B_\text{est}(t+\dd t) = B_\text{est}(t) + \gamma \partial_B l_t$ with $B_\text{est}(0)=0.5$ and $\gamma = 0.5 \dd t$.
(d) Expectations of $\sigma_x/2$ and $\sigma_z/2$ using true $\rho_t$ and the estimated $\rho_\text{est}$. The agreement is good for $t>200$.
}
\label{fig:online_parameter_est}
\end{figure}

In this section we consider the problem of real-time parameter estimation.
The goal of the magnetometer [Eq.~\eqref{eqn:SDE_outcome}] is to estimate $B$ from the measurement outcomes $\{ \dd Y_t \}$. The typical initial state is the non-entangled spin-coherent state $\rho_0 =  | \psi_0 \rangle  \langle \psi_0 |$ where $| \psi_0 \rangle$ is the maximum-eigenvalue state of $\hat{J}_x$ (i.e., $ \hat{J}_x| \psi_0 \rangle = \frac{N}{2} | \psi_0 \rangle$). The standard parameter estimation is based on maximum likelihood \cite{PhysRevA.87.032115, Albarelli_2017}: the estimated $B_\text{est}$ is determined by the $B$ value that gives the maximum probability of the quantum trajectory specified by  $\{ \dd Y_t \}$, and the implementations typically require scanning or sampling the parameter-to-estimate over a predefined range \cite{PhysRevA.87.032115} [see also Fig.~\ref{fig:online_parameter_est}(a)]. In the large-$N$ limit, quantum dynamics can be simplified by the Gaussian approximation
(i.e., $[\hat{J}_y, \hat{J}_z] = i \hat{J}_x \approx i \langle J_x \rangle(t)$ whose validity originates from the chosen initial state), and $B$ can be estimated in real time (without scanning) using the Quantum Kalman filter (i.e., field $B$ is included in the dynamical equation)  \cite{PhysRevLett.91.250801, PhysRevA.69.032109}.

Here we consider the real-time parameter estimation with the full quantum dynamics. Real-time parameter estimation requires updating the quantum state $\rho_\text{est}$ and the field $B_\text{est}$ simultaneously. One natural argument against this scenario is that the quantum state evolved according to the estimated $B_\text{est}$ is different from that according to $B_\text{true}$, %($B_\text{est} \neq B_\text{true}$ during the early evolution) and % so the quantum state evolved accordingly  %evolves according to the wrong field and cannot be correct;
and it is {\em not} obvious if one can ever recover the correct state and therefore the field $B_\text{true}$ within a single evolution.
On the other hand, the previous analysis shows that the asymptotic state depends only on the measurement outcomes $\{ \dd Y_t \}$, not on the initial state. If $B_\text{est}$ is ''reasonably`` correct over certain time window, one can recover the quantum state and thus the correct $B$. We shall demonstrate that, the real-time parameter estimation is possible if the sign and the amplitude range of $B$ are known.

%Criterion (rule) for B-update: time-local log-likelihood Trap in local minimum; but this problem only has two local minima. Technical contribution: Stochastic DE for (d log-likelihood/dB)

%{app:SDE_LL}

\subsection{Problem setup and SDE of likelihood function}

The problem setup for real-time $B$-estimation is as follows: fixing $-B = B_\text{true}$ in Eq.~\eqref{eqn:SDE_outcome} (the minus sign is due to our convention) and generating a corresponding measurement outcomes $\{ \dd Y_t \}_{B_\text{true}}$, device the procedure to update $B_\text{est}$ and $\rho_\text{est}$ such that $B_\text{est}$ approaches $B_\text{true}$ in the long-time limit.

Let us review the SDE for the likelihood function first derived in Ref.~\cite{PhysRevA.87.032115}. %
%\subsection{SDE for $\partial_B \log L_t$ and $\partial^2_B \log L_t$}
The likelihood function $L_t(B)$ is given by $L_t(B) = \text{tr} [ \tilde{\rho}_t]$ where $\tilde{\rho}_t$ is the un-normalized DM.
%\beq
%\begin{aligned}
%\Omega (\dd Y_t) &=\mathbb{I} - i B \hat{J}_y \,\dd t - \frac{1}{2} \hat{J}_z^2 \, \dd t + \hat{J}_z \dd Y_t, \end{aligned}  \label{eqn:Kraus_unnormalized}
%\eeq
%This implies $\dd Y_t = \text{tr}[ \mathcal{M}(\rho_t) ] dt + dW_t$ with $dW_t$  a Wiener increment with zero mean and variance $dt$.
Using Eq.~\eqref{eqn:Kraus_unnormalized}, the SDE for un-normalized $\tilde{\rho}_t$ is given by
\beq
\begin{aligned}
\dd \tilde{\rho}_{t}  &\equiv \tilde{\rho}_{t+dt} - \tilde{\rho}_{t} =
\Omega (\dd Y_t) \tilde{\rho}_{t}  \Omega^\dagger (\dd Y_t) - \tilde{\rho}_{t} \\
&= \underbrace{
\left( -i[B \hat{J}_y, \tilde{\rho}_{t}] - \frac{1}{2} \{ \hat{J}_z^2, \tilde{\rho}_{t} \} + \hat{J}_z \tilde{\rho}_{t} \hat{J}_z \right) }_{\text{traceless}} \dd t+
\underbrace{ \left( \hat{J}_z \tilde{\rho}_{t} + \tilde{\rho}_{t} \hat{J}_z \right) }_{\equiv \mathcal{M}(\tilde{\rho}_t)} \dd Y_t.
\end{aligned}
\label{eqn:diffusion_DM_unnormalized}
\eeq
Taking the trace of \eqref{eqn:diffusion_DM_unnormalized} one gets the SDE of the likelihood $L_t$
\begin{align}
\dd L_t &= \text{tr}(\tilde{\rho}_{t+dt} ) - \text{tr}(\tilde{\rho}_t) = \text{tr} \left[\mathcal{M}( \tilde{\rho}_t) \right] \dd Y_t =
\text{tr} \left[\mathcal{M}( {\rho}_t) \right] L_t \dd Y_t. %\\
%%%%%%%%%%%%% \tilde{\rho}_t &= L_t {\rho}_t \Leftrightarrow {\rho}_t = \frac{\tilde{\rho}_t }{L_t}.
\label{eqn:likelihood_02}
\end{align}
Numerically it is easier to work with the log-likelihood function $l_t(B) = \log L_t(B)$ as $L_t(B)$ becomes exponentially small away from its maximum value. Using $\dd l_t = \frac{\dd L_t}{L_t} - \frac{1}{2} \frac{(\dd L_t)^2}{L^2_t} + o(\dd t)$ and $(\dd Y_t)^2=\dd t$ one gets
\beq
\begin{aligned}
\dd l_t &= \text{Tr} [ \mathcal{M}(\rho_t )]\left\{ \dd Y_t - \frac{1}{2} \text{Tr} [  \mathcal{M}(\rho_t ) ]\dd t \right\}, \\
%%%
l_t &= \sum_{t'<t} \dd l_t (t').
\end{aligned}
\label{eqn:dl_t_correct}
\eeq
Eq.~\eqref{eqn:dl_t_correct} allows us to estimate $B$ using maximum likelihood. In Fig.~\ref{fig:online_parameter_est}(a) we show $L_t(B)$ for $B \in [-1.5, 1.5]$ with $N=1$ (single qubit) and $\{ \dd Y_t \}_{ B_\text{true}=1}$. The inset of Fig.~\ref{fig:online_parameter_est}(a) gives the $l_t(B)$ at $t=400$. It is seen that there are two maxima: the global one at $B=B_\text{true}$; the local one at $B=-B_\text{true}$. Around $t=200$ it converges to the correct $B=B_\text{true}$. The most time consuming part of maximum likelihood method is the parameter scanning, and we proceed to the real-time problem in the next subsection.

\subsection{Parameter updating based on local maximum likelihood}

%${\dd Y_t}$ is generated by  with a known $-B = B_\text{true}=1$
The real-time problem aims to use a single calculation, instead of $B$-scanning, for parameter estimation. It requires evolving the quantum state $\rho_\text{est}$ and the field $B_\text{est}(t)$ at the same time. $\rho_\text{est}$ is evolved according to its SME with a time-dependent $B_\text{est}(t)$. The most natural choice of $B$-updating is to maximize the change of log-likelihood by following the gradient $\partial_B (\dd l_t) \equiv \dd l_t^B$. The SDE for $\dd l_t^B$ is derived in Eq.~\eqref{eqn:dl^B_dt_SDE} of Appendix \ref{app:SDE_LL} and is numerically tested by comparing to the finite-difference approximation [see Fig.~\ref{fig:online_parameter_est}(b)]. Since $B$-updating is based on the gradient search but $l_t(B)$ actually has two maxima at $B=\pm B_\text{true}$ [Fig.~\ref{fig:online_parameter_est}(a)], we need a prior knowledge of the sign of $B$-field so that $B_\text{est}$ converges to the global maximum. We assume that the sign and the amplitude bound of $B$ are known; this can be experimentally implemented by applying a large DC offset field.

The overall dynamics of $\rho_\text{est}(t)$ and $B_\text{est}(t)$ are
\begin{subequations}
\begin{align}
& \dd \rho_\text{est} =   -i \big[ B_\text{est} (t) \hat{J}_y, \rho_\text{est} \big] \dd t + \mathcal{D}[\hat{J}_z] \rho_\text{est} \, \dd t \nonumber \\
& \,\,\,\, + \left\{ \rho_\text{est} \hat{J}_z + \hat{J}_z \rho_\text{est} -  2 \ \text{tr}[ \rho_\text{est} \hat{J}_z]   \rho_\text{est} \right\} \cdot \left\{ \dd Y_t - 2 \ \text{tr}[ \rho_\text{est} \hat{J}_z]  \dd t  \right\},
\label{eqn:rho_est} \\
%%%%%%%%%%
& B_\text{est} (t + \dd t) =
B_\text{est} (t) + \gamma \cdot \dd l^B_t,
\label{eqn:B_est}
\end{align}
\label{eqn:drho_SDE_est}
\end{subequations}
with $B_\text{est} (t)$ restricted to $[0, B_\text{max}]$: if $B_\text{est} (t)< 0$ then $B_\text{est} (t)= 0$; if $B_\text{est} (t) > B_\text{max}$ then $B_\text{est} (t)= B_\text{max}$. $\gamma $ is the updating rate, which is empirically found to be of the order of $\dd t$.
In Fig.~\ref{fig:online_parameter_est}(c) we show the $B_\text{est}(t)$ for three realizations of $\{ \dd Y_t\}_{B_\text{true}=1}$. They all approach but fluctuate around $B=B_\text{true}$ once $t \gtrsim 200$. Some averaging procedure can be applied to obtain a smoother $B_\text{est} (t)$.  Fig.~\ref{fig:online_parameter_est}(d) further compares $\text{tr}[ \rho_t \frac{\sigma_i}{2} ]$ with $\text{tr}[ \rho_\text{est} \frac{\sigma_i}{2} ]$ with $i=x,z$; the agreement is also good. We have tested up to $N=40$ and the qualitative behavior is the same.

As a brief summary, the real-time parameter estimation based on local maximum likelihood appears to work once the sign and the amplitude of $B$-field are known. It may be a bit surprising that the error of estimated quantum state, caused mainly by the inaccurate $B_\text{est}(t)$, accumulated in the early evolution can be compensated within a single calculation [Eqs.~\eqref{eqn:drho_SDE_est}]. According to the analysis in Section \ref{sec:expt_implication}, as far as  $B_\text{est}(t) \approx B_\text{true}$ over a time interval, {\em any} quantum state evolving using the same $\{ \dd Y_t \}$ converges to the same state. In other words, the intentional compensation for state error is not needed because the long-time quantum state is dominantly determined by $\{ \dd Y_t \}$.
The key problem is therefore if the local maximum likelihood is sufficient to have $B_\text{est}(t) \approx B_\text{true}$. What we demonstrate is that by restricting $B_\text{est}$ in the domain where $l_t(B)$ only has one maximum, the gradient based update [Eq.~\eqref{eqn:B_est}] does give a $B_\text{est}(t)$ that is fluctuating around $B_\text{true}$.

\section{Conclusion }

To conclude, we investigate the asymptotic behavior of quantum trajectories under weak continuous measurements for magnetometer without decoherence. %Two general behaviors are identified: (i) the system always becomes pure asymptotically; (ii)
We numerically confirm that: given one realization the quantum trajectories starting from different initial states converge to the same realization-specific pure state.
For single-qubit systems we are able to prove its generality. Specifically we first apply the probability theory to show that the quantum trajectory always becomes pure asymptotically. Confined in subspace of pure states we derive a non-linear stochastic equation; by solving the corresponding time-independent Fokker-Planck equation we show that asymptotically any two quantum trajectories coincide.
The numerical simulations strongly support its validity for the system of multiple qubits.
Related to experiments, we show (without a complete proof even for the single-qubit case) that if one starts with a wrong initial state and updates the system conditioned on the experimental outcomes, the asymptotic quantum trajectory eventually coincides with the one starting with the correct initial state.
%Namely, the asymptotic pure state depends only on the measurement outcomes, not on the initial state.
The long-time behavior being insensitive to the initial state suggests that the real-time parameter estimation is possible: as far as the estimated parameter is correct over a certain period of time, the errors in the quantum states becomes negligible. We illustrate it by explicitly proposing an stochastic equation that simultaneously updates the estimated quantum state and parameter based on the local maximum likelihood; a reasonable result is seen.
Looking forward, we believe the problem considered here can be a useful and experimentally relevant test bed for many interesting physics such as quantum decoherence and measurement efficiency.
%With the analysis, we explore the real-time parameter estimation and find that the local maximum likelihood with some constraints does give a very reasonable result.

%with a wrong initial state 

%Our results imply that when a quantum task involves the continuous measurement, the initial state is not and cannot be critical asymptotically. For example, for quantum tomography the estimation of the initial state has to be done early before the state becomes insensitive to where it starts; for quantum metrology the state preparation may not be very critical using weak measurements and identifying the optimal measuring time can be an interesting problem.
%Although the problem considered here is specific,

\section*{Acknowledgment} 
CL thanks Andrew Millis (Columbia University and Flatiron Institute) for a very helpful discussion.

%\section{Conclusion} 

%%%%%%%%%%%%%%%%%%%%%%%%%%%%%%%%%%%%%%%%
\appendix 

%\section{SDE for polar coordinate } \label{app:polar_equation}
\section{Convergence to the stationary distribution} \label{app:stationary}

\subsection{General derivation}
We outline the derivation given in Chapter 6.1 of Ref.~\cite{book_FP_Risken}. Consider two distribution function $W_1 (\mathbf{x},t)$ and $W_2(\mathbf{x},t)$ that satisfy the same FP equation $\partial_t W = L_\text{FP} W$. We want to prove that $W_1 (\mathbf{x},t)$ and $W_2(\mathbf{x},t)$ asymptotically converge to the same distribution. If the stationary distribution exists, then they converge to the stationary distribution. The key is to consider the time evolution of the Kullback-Leibler (KL) divergence
\beq 
H(t) = \int \dd \mathbf{x} \ W_1 \ln \frac{W_1}{W_2} 
=  \int \dd \mathbf{x} \ W_1 \ln R,
\eeq 
where $R = \frac{W_1}{W_2} > 0 $ is the ratio of two probability distributions. By construction KL divergence is non-negative, i.e., $H(t) \geq 0$.  After some manipulations \cite{book_FP_Risken} one arrives  
\beq 
\begin{aligned} 
 \dot{H}(t) &= - \int \dd \mathbf{x} \bigg[ W_1 D_{ij} \frac{\partial\ln  R}{\partial x_j}   \frac{\partial \ln R}{\partial x_i}  \bigg] \leq 0.
\end{aligned}
\label{eqn:Hdot->R1}
\eeq 
where $D_{ij}$ is the diffusion coefficient(s). 

%[Do I need to use continuous condition??]

For a positive definite $D_{ij}$, ${H}(t)$ keeps decreasing once $\frac{\partial \ln R}{\partial x_i} \neq 0$.  
Because $H(t)$ is bounded from below, $\dot{H}$ has to approach zero; $\dot{H}=0$ implies $\ln R$ (and thus $R$) becomes independent of $\mathbf{x}$. Because of the normalization, $R$ must be equal to 1 and $H(t)$ reaches its lower bound $H = 0$. Thus any two solutions $W_1$ and $W_2$ must coincide for large times. The arguments are summarized in Eq.~\eqref{eqn:flow_}:
\begin{subequations}
 \begin{align} 
 %& \text{If }D_{ij}>0 \Rightarrow \dot{H}(t) \leq 0 \text{ when } \frac{\partial \ln R}{\partial x_i} \neq 0 \\
 %%%%
  & H(t) \geq 0 \text{ and } \dot{H}(t) \leq 0 \Rightarrow \lim_{t\rightarrow \infty} \dot{H}(t) \rightarrow 0  \\
  & \dot{H}(t) = 0  \text{ and } D_{ij}>0 \underset{ \text{\eqref{eqn:Hdot->R1}} }{ \Rightarrow  }
  \frac{\partial \ln R}{\partial x_i} = 0 \nonumber \\
  %%%
  & \ \ \Rightarrow R = \frac{W_1}{W_2} = 1 \text{ and } H=0. % \text{ asymptotically}.
  \label{eqn:R1}  
  %& \int \dd \mathbf{x} \ W_i (\mathbf{x}) = 1 \Rightarrow R = 1 
  %\Rightarrow \lim_{t\rightarrow \infty} H(t) \rightarrow 0 \text{ and }
 \end{align}
\label{eqn:flow_}
\end{subequations}
If the drift and diffusion coefficients are time independent, a stationary solution $ L_\text{FP} W_\text{sta} = 0$ may exist. It then follows from $\dot{H}=0$ (i.e., $R=1$ for all $\mathbf{x}$) that this solution is {\em unique} and that all other $W$'s asymptotically agree with it.  
We remark that in the derivations $W$ is assumed to be non-zero for the entire domain. 
$\delta$-function is allowed if we regard it as a limit of Gaussian distribution. 

%It should be noted that this result is valid if $D_{ij}$ is positive  definite everywhere and if the drift coefficient has no singularities, though the time $T$ may be very large. 

\subsection{Applications to Eq.~\eqref{eqn:FP_theta} and Eq.~\eqref{eqn:FP_theta_theta}}

The key for reaching the unique asymptotic stationary distribution is that when $D_{ij}>0$, $\dot{H}(t)=0$ implies $R(x)=1$ for the entire domain of interest [Eq.~\eqref{eqn:Hdot->R1} and \eqref{eqn:R1}]. For FP equations considered in the main text, $D_{ij}$ is only non-negative definite and we discuss how to get around it using the continuity and periodicity.

For the single-variable SDE Eq.~\eqref{eqn:FP_theta}, the diffusion coefficient $D(\theta) = \frac{1}{2} \cos^2 \theta$ which are zero when $\theta = \pm \frac{\pi}{2}$ so that $\dot{H}=0$ implies $R$ ($= \frac{W_1(\theta, t)}{ W_2(\theta, t) }$) is not necessarily one at $\theta = \pm \pi/2$. However if we impose the continuity conditions for $W$'s and thus $R$, then $R(\pm \frac{\pi}{2} + \varepsilon) = 1$ implies $R(\pm \frac{\pi}{2})=1$. Therefore for Eq.~\eqref{eqn:FP_theta}, as far as $P^{(1)}_\text{sta}(\theta)$ is non-zero for the entire domain it is unique and will be asymptotically reached for any initial distributions. 

For the two-variable SDE Eq.~\eqref{eqn:FP_theta_theta},  the diffusion matrix
\beq 
D(\theta, \bar{\theta}) = \frac{1}{2} 
\begin{bmatrix} \cos^2 \theta &  \cos \theta \cos \bar{\theta} \\
 \cos \theta \cos \bar{\theta} & \cos^2 \bar{\theta} \end{bmatrix}
\label{eqn:D_2variable}
\eeq 
which is positive semidefinite: it has one non-negative eigenvalue $\frac{1}{2} (\cos^2 \theta + \cos^2 \bar{\theta}) $ with the eigenvector $[\cos \theta, \cos \bar{\theta}]^T$ and a zero eigenvalue with the eigenvector $[-\cos \bar{\theta}, \cos \theta]^T$. Define $R(\theta, \bar{\theta}, t) = \frac{W_1(\theta, \bar{\theta}, t)}{ W_2(\theta, \bar{\theta}, t) }$, Eq.~\eqref{eqn:Hdot->R1} gives 
\beq 
\begin{aligned} 
 \dot{H}(t) &= -\frac{1}{2} \int_{-\pi}^\pi \int_{-\pi}^\pi  \dd \theta \dd \bar{\theta} \ W_1   \bigg[ \cos \theta \ \frac{\partial\ln  R}{\partial \theta}  +  \cos \bar{\theta} \ \frac{\partial  \ln  R}{\partial \bar{\theta}} \bigg]^2 \leq 0.
\end{aligned}
\label{eqn:Hdot->R1_theta2}
\eeq  
From Eq.~\eqref{eqn:Hdot->R1_theta2}, $\dot{H}=0$ only implies $\cos \theta \ \frac{\partial\ln  R}{\partial \theta}  +  \cos \bar{\theta} \ \frac{\partial  \ln  R}{\partial \bar{\theta}} =0$. %Certainly constant $R$ is a solution. 
Let us construct the non-constant solution that is continuous and period. Using separation of variable we express $\ln  R \equiv F(\theta) \bar{F}( \bar{\theta} )$ to get 
\beq 
\begin{aligned}
& F'(\theta) \cos (\theta) \bar{F}(\bar{\theta}) 
+ F(\theta)  \bar{F}'(\bar{\theta})  \cos (\bar{\theta}) = 0 \\
%%%%%%%%
\Rightarrow \ & 
\frac{ F'(\theta) \cos (\theta) }{ F(\theta)  } 
+ \frac{ \bar{F}'(\bar{\theta})  \cos (\bar{\theta}) }{\bar{F}(\bar{\theta}) } = 0 \\
%%%%%%%
\Rightarrow & 
\frac{ F'(\theta) \cos (\theta) }{ F(\theta)  } = -
\frac{ \bar{F}'(\bar{\theta})  \cos (\bar{\theta}) }{\bar{F}(\bar{\theta}) } = a,
\end{aligned}
\label{eqn:Sepa_variable}
\eeq 
with $a$ a real-valued constant. $a=0$ corresponds to a constant solution. When $a \neq 0$ we can write 
\beq 
\ln \frac{ F(\theta) }{ F(-\pi) }= + \int_{- \pi}^\theta \dd \theta' \frac{a}{ \cos \theta }.
\label{eqn:diverge_F}
\eeq 
$F(\theta)$ is discontinuous at $\theta = \pm \frac{\pi}{2}$. Alternatively, substitute the Fourier series $F(\theta) = \sum_{n=-\infty}^{\infty} f_n e^{i n \theta}$ (with the real-valued constraint $f_{-n} = f_n^*$) into Eq.~\eqref{eqn:Sepa_variable} leads to the recursion relation $a f_n = i \big[\frac{n-1}{2} f_{n-1} + \frac{n+1}{2} f_{n+1}  \big]$ that does not have a real-valued solution. To recap, requiring $\ln R$ to be real and periodic by direct integration leads to a discontinuity [Eq.~\eqref{eqn:diverge_F}]; requiring $\ln R$ to be periodic and continuous using Fourier expansion leads to a complex expression.
Therefore if we impose that $W(\theta, \bar{\theta}, t)$ (and thus $\ln R$) is a continuous,  periodic and real-valued function in $(\theta, \bar{\theta})$, then $a=0$ in Eq.~\eqref{eqn:Sepa_variable} and $R$ has to be one when $\dot{H}=0$. We thus conclude that for Eq.~\eqref{eqn:FP_theta_theta},  any initial distributions eventually converge to the stationary distribution $P^{2}_\text{sta} (\theta , \bar{\theta}) = P^{(1)}_\text{sta}(\theta) \delta(\theta - \bar{\theta})$.

We wish to point out again that the proof requires the existence of a non-zero probability distribution over the entire domain; allowing the $\delta$-function (as the limit of Gaussian distribution) is reasonable but can only be regarded as an assumption.

\subsection{Stationary solution for Eq.~\eqref{eqn:FP_theta_theta}} \label{app:2v_sta}

We show that $P^{(2)}_\text{sta} = P^{(1)}_\text{sta} (\theta) \delta(\theta -\bar{\theta})$ is the solution of $L_\text{FP}(\theta, \bar{\theta}) P^{(2)}_\text{sta} = 0$. To simplify the notation we define $b(\theta) = B + \frac{1}{4} \sin (2 \theta) $ and rewrite the equation \eqref{eqn:FP_theta_stationary} as 
\beq 
0 = - \frac{\partial}{\partial \theta} \big[  b(\theta) P^{(1)}_\text{sta} (\theta) \big] 
+ \frac{1}{2} \frac{\partial^2}{\partial \theta^2} \big[ \cos^2 (\theta) \ P^{(1)}_\text{sta} (\theta) \big].
\label{eqn:P1_sta_rep}
\eeq 
As the $P^{(2)}_\text{sta}$ involves $\delta$-function, we use the weak form: for an arbitrary periodic function $F(\theta, \bar{\theta}) = F(\theta + 2 m \pi, \bar{\theta} + 2 \bar{m} \pi)$ ($m$, $\bar{m}$ being integers), 
\begin{equation}
\begin{aligned} 
0 &= \int_{-\pi}^\pi \int_{-\pi}^\pi \ \dd \theta \ \dd \bar{\theta} \ F(\theta, \bar{\theta}) \cdot  \\ 
 %%%%%%%%%%%%%%%%
&  \begin{pmatrix*}[l] 
- \frac{\partial}{\partial \theta} \big[b(\theta) P^{(1)}_\text{sta} (\theta) \delta(\theta -\bar{\theta})  \big] & \text{(i-1)} \\
%%%% 
- \frac{\partial}{\partial \bar{\theta} } \big[b(\bar{\theta}) P^{(1)}_\text{sta} (\theta) \delta(\theta -\bar{\theta})  \big] & \text{(ii-1)} \\
%%%
 + \frac{1}{2} \frac{\partial^2}{\partial \theta^2} \big[ \cos^2( \theta) \ P^{(1)}_\text{sta} (\theta) \delta(\theta -\bar{\theta}) \big]  & \text{(i-2)} \\
+ \frac{1}{2} \frac{\partial^2}{\partial \theta \partial \bar{\theta}}  
 \big[ \cos(\theta) \cos(\bar{\theta})  P^{(1)}_\text{sta} (\theta) \delta(\theta -\bar{\theta}) \big]  & \text{(i-3)} \\
 +  \frac{1}{2} \frac{\partial^2}{\partial \theta \partial \bar{\theta}}  
 \big[ \cos(\theta) \cos(\bar{\theta})  P^{(1)}_\text{sta} (\theta) \delta(\theta -\bar{\theta}) \big] & \text{(ii-2)} \\ 
+ \frac{1}{2} \frac{\partial^2}{\partial \bar{\theta}^2} \big[ \cos^2( \bar{\theta} ) \ P^{(1)}_\text{sta} (\theta) \delta(\theta -\bar{\theta}) \big]     & \text{(ii-3)}  
\end{pmatrix*} .
\end{aligned} 
\label{eqn:sol_weak}
\end{equation}  
In Eq.~\eqref{eqn:sol_weak} it turns out that (i-1) + (i-2) + (i-3) = 0 and (ii-1) + (ii-2) + (ii-3) = 0; we only show the former. Because all functions are periodic, when using integration by part the boundary terms are zero. To avoid any confusions we explicitly note that $F(\theta, \bar{\theta})$ has two arguments; the subscript indicates the argument for partial differentiation: $F_\theta = \frac{\partial}{\partial \theta} F(\theta, \bar{\theta})$, $F_{\bar{\theta}} = \frac{\partial}{\partial \bar{\theta} } F(\theta, \bar{\theta})$, $F_{\theta \bar{\theta}} = \frac{\partial^2}{\partial \theta \partial \bar{\theta} } F(\theta, \bar{\theta})$... etc. (i-1) of Eq.~\eqref{eqn:sol_weak} is 
\beq 
\begin{aligned}
A_\text{i-1} &=  \int  \dd \theta \ \dd \bar{\theta} \  \big[b(\theta) P^{(1)}_\text{sta} (\theta) \delta(\theta -\bar{\theta})  \big] F_\theta(\theta, \bar{\theta}) \\
&= \int  \dd \theta \ b(\theta) P^{(1)}_\text{sta} (\theta)  F_\theta(\theta, \bar{\theta} = \theta)  \\
&= - \int  \dd \theta \ \frac{\partial}{\partial \theta}\big[ b(\theta) P^{(1)}_\text{sta} (\theta)  \big] G(\theta), \text{ where } \\
& \frac{\dd}{\dd \theta} G(\theta) \equiv F_\theta(\theta,  \bar{\theta}=\theta).
\end{aligned}
\label{eqm:i-1}
\eeq 
$G(\theta)$ is a single-argument function and its second derivative is given by $\frac{\dd^2}{\dd \theta^2} G(\theta) = F_{\theta \theta} (\theta,  \bar{\theta}=\theta) + F_{\theta \bar{\theta} } (\theta,  \bar{\theta}= \theta)$. 
(i-2) of Eq.~\eqref{eqn:sol_weak} is 
\beq 
\begin{aligned}
A_\text{i-2} &=  \frac{1}{2} \int  \dd \theta \ \dd \bar{\theta} \  \big[ \cos^2( \theta) \ P^{(1)}_\text{sta} (\theta) \delta(\theta -\bar{\theta}) \big] F_{\theta \theta} (\theta, \bar{\theta}) \\
&=  \frac{1}{2}  \int  \dd \theta \  \cos^2( \theta)  P^{(1)}_\text{sta} (\theta)  \underbrace{ F_{\theta \theta} (\theta, \bar{\theta} = \theta) }_{ = \frac{\dd^2}{\dd \theta^2} G(\theta) - F_{\theta \bar{\theta} } (\theta,  \bar{\theta}= \theta) } \\ 
%%%%%%%
&= \frac{1}{2} \int  \dd \theta \ \frac{\partial^2}{\partial \theta^2}\big[  \cos^2( \theta)  P^{(1)}_\text{sta} (\theta)  \big] G(\theta) \\
&\ \ -  \frac{1}{2}  \int  \dd \theta \  \cos^2( \theta)  P^{(1)}_\text{sta} (\theta) F_{\theta \bar{\theta} } (\theta,  \bar{\theta}= \theta) .
\end{aligned}
\label{eqm:i-2}
\eeq 
(i-3) of Eq.~\eqref{eqn:sol_weak} is 
\beq 
\begin{aligned}
A_\text{i-3} &=  \frac{1}{2} \int  \dd \theta \ \dd \bar{\theta} \  \big[ \cos( \theta) \cos( \bar{\theta} ) \ P^{(1)}_\text{sta} (\theta) \delta(\theta -\bar{\theta}) \big] F_{\theta \bar{\theta} } (\theta, \bar{\theta}) \\
&=  \frac{1}{2}  \int  \dd \theta \  \cos^2( \theta)  P^{(1)}_\text{sta} (\theta)  F_{\theta \theta} (\theta, \bar{\theta} = \theta).
\end{aligned}
\label{eqm:i-3}
\eeq  
The sum of Eq.~\eqref{eqm:i-1}, \eqref{eqm:i-2}, \eqref{eqm:i-3} is zero. Particularly the sum of Eq.~\eqref{eqm:i-1}  and the first term of \eqref{eqm:i-2} is zero due to Eq.~\eqref{eqn:P1_sta_rep}; the second term of \eqref{eqm:i-2} and \eqref{eqm:i-3} cancel each other. The same procedure leads to (ii-1) + (ii-2) + (ii-3) = 0 in Eq.~\eqref{eqn:sol_weak}. This completes the proof.

\subsection{About experimental implication} \label{app:expt_implication}

Here we consider single-qubit $\rho^{(0)}(t)$, $\rho^{(2)}(t)$ in Section \ref{sec:expt_implication}. Assuming both states are pure with zero $\rho_y$ component so one uses $\bar{\theta}$ for $\rho^{(0)}$ and ${\theta}$ for $\rho^{(2)}$, the coupled the SDE is
\begin{equation}
\begin{aligned} 
 %%%%%%%
 \dd  \begin{bmatrix} \theta \\ \bar{\theta} \end{bmatrix} &=  \begin{bmatrix} \tilde{b}(\theta) \\ b(\bar{\theta})     \end{bmatrix}  \dd t + 
 \begin{bmatrix}  \cos \theta \\ \cos \bar{\theta} \end{bmatrix}  \dd W_t, \\
 %%%%%%%
 \text{where } &  \tilde{b}(\theta) = b(\theta) + \cos(\theta) [ \cos(\bar{\theta}) - \cos(\theta)  ]
 \label{eqn:theta_2_expt}
\end{aligned} 
\end{equation}   
Compared to Eq.~\eqref{eqn:1D_couple_origin_3}, the only difference is the drift term of $\theta$, i.e., $b(\theta) \rightarrow \tilde{b}(\theta)$. The stationary FP equation from Eq.~\eqref{eqn:theta_2_expt} is  
\begin{equation}
\begin{aligned} 
0 &=
 - \frac{\partial}{\partial \theta} \big[ \tilde{b}(\theta) P^{(2)}  \big]
 - \frac{\partial}{\partial \bar{\theta} } \big[{b}(\bar{\theta} ) P^{(2)}  \big] \\ 
 %%%%%%%%%%%%%%%%
&+ \frac{1}{2} \begin{pmatrix*}[l]
 + \frac{\partial^2}{\partial \theta^2} \big[ \cos^2( \theta) \ P^{(2)}\big] \\
+ 2 \frac{\partial^2}{\partial \theta \partial \bar{\theta}}  
 \big[ \cos(\theta) \cos(\bar{\theta})  P^{(2)}\big]
\\ 
+ \frac{\partial^2}{\partial \bar{\theta}^2} \big[ \cos^2( \bar{\theta} ) \ P^{(2)}\big]       
\end{pmatrix*} . 
\end{aligned} 
\label{eqn:FP_theta_theta_expt}
\end{equation}  
To see that $P^{(2)}_\text{sta} =  P^{(1)}_\text{sta}(\theta) \delta(\theta - \bar{\theta})$ is a also a solution of Eq.~\eqref{eqn:FP_theta_theta_expt}, we note that replacing $b(\theta)$ by  $\tilde{b}(\theta)$ in the first equality of Eq.~\eqref{eqm:i-1} does not change the answer thanks to the $\delta(\theta - \bar{\theta})$ term. The remaining steps are identical to those in the Appendix \ref{app:2v_sta}. We cannot prove the asymptotic purity of $\rho^{(2)}$ and have to assume it, but Eqs.~\eqref{eqn:rho_y2} and \eqref{eqn:purestate_explicit_02} do guarantee that once $\rho^{(2)}$ is pure it remains pure.

%%%%%%%%%%%%
\section{Continued fraction} \label{app:CF} 

Two lists of numbers $\mathbf{a} = [a_1, a_2, \cdots, ]$, $\mathbf{b} = [b_1, b_2, \cdots, ]$ define the continued fraction: 
\beq 
x= \cfrac{a_{1}}{ b_{1} + \cfrac{a_{2}}{b_{2}+ \cfrac{a_{3}}{b_{3}+ \cdots  } } }    
\equiv \frac{ a_1 | }{ |b_1 } + \frac{ a_2 | }{ |b_2 } + \frac{ a_3 | }{ |b_3 } + \cdots
\eeq 
The second expression is referred to as the Pringsheim notation. Some terminologies are provided \cite{book_CF_Wall}. $\mathbf{a}$ is the list of {\em partial numerator}; $a_p$ is $p$th partial numerator. $\mathbf{b}$ is the list of {\em partial denominator}; $b_p$ is $p$th partial denominator. $\frac{a_p}{b_p}$ is called $p$th {\em partial quotient}. The truncated continued fraction
\beq 
\begin{aligned}
x_n &= \frac{ a_1 | }{ |b_1 } + \frac{ a_2 | }{ |b_2 } + \cdots + \frac{ a_n | }{ |b_n } 
\end{aligned} 
\label{eqn:approximant}
\eeq  
is called $n$th {\em approximant} or $n$th {\em convergent} of $x$ and $x = \lim_{n \rightarrow \infty} x_n$ if the limit exists.

We now prove Eq.~\eqref{eqn:Pstat_B0}. For $B=0$, Eq.~\eqref{eqn:iteration} becomes
\beq  
\begin{aligned}
& \underbrace{  1  }_{ {Q}_m } a_m + \underbrace{  \frac{1}{2} \frac{m-1}{m}  }_{ {Q}^+_m }  a_{m+2} + \underbrace{ \frac{1}{2} \frac{m+1}{m} }_{ {Q}^-_m } a_{m-2}= 0. \\
%%%%%%%%%%%%%%%%
\Rightarrow &  
\mathbf{ {a} }_m = [ -{Q}^-_{m+2}, -{Q}^+_{m+2} {Q}^-_{m+4}, -{Q}^+_{m+4} {Q}^-_{m+6}, \cdots  ]  \\
&= \bigg[ -\frac{1}{2} \frac{m+3}{m+2}, - \frac{ (m+1)(m+5) }{4(m+2)(m+4)}, - \frac{ (m+3)(m+7) }{4(m+4)(m+6)}, - \frac{ (m+5)(m+9) }{4(m+6)(m+8)}  \cdots \bigg]\\
%%%%%%%%%%%%%%%%%%%
&\mathbf{ {b} }_m  = [ {Q}_{m+2}, {Q}_{m+4}, {Q}_{m+6}, \cdots  ] 
= [ 1, 1, 1, \cdots  ]
\end{aligned}
\label{eqn:iteration_B0} 
\eeq  
The goal is to show that $S_{2m} = -1$ so that $a_{2m} = \frac{ (-1)^m }{2 \pi}$. To do so we use $Q_m=1$ to get the recursive relation   
$ \frac{Q^-_{m+2} }{ S_{m} } + 1 + Q^+_{m+2} S_{m+2} = 0
$.
Using $Q^{\pm}_{m} = \frac{1}{2} \frac{ m \mp 1}{m}$, we find that {\em if} $S_{m+2} = -1$ then $S_{m} = -1$ and vise versa. Therefore if one finds  $S_{2m} = -1$ for an integer $m$, then $S_{2m} = -1$ for all integers $m$.  Taking $m \rightarrow \infty$, we get 
\beq 
S_\infty %= \cfrac{  - 1/2 }{ 1  - \cfrac{ 1/4 }{ 1 
  %- \cfrac{1/4 }{1 - \cdots}   }  }  
  = \cfrac{  - 1/2 }{ 1  + \frac{1}{2} \times  \cfrac{ -1/2 }{ 1 
  - \cfrac{1/4 }{1 - \cdots}   }  }  
  = \cfrac{  - 1/2 }{ 1  + \frac{1}{2} S_\infty } 
  \Rightarrow S_\infty = -1. 
\eeq 
This completes the proof.

\section{Lyapunov analysis (Stability)} \label{app:Lyapunov}

Stability is to address if $X_t$ is bounded (stable)  or approaches a stable point (asymptotically stable) as $t \rightarrow \infty$. The stable point is assumed to be $x_0$. One way to address this question is to introduce a Lyapunov function $v(t,x)$, and analyze the behavior of an ``alternative'' stochastic process $V_t = v(t, X_t)$ that is adapted to $X_t$. 

For the SDE of Eq.~\eqref{eqn:SDE_example}, a ''generator operator`` (an operator generated from a given SDE) acting on $t$ and $x$   is defined as
\beq 
L = \frac{\partial}{\partial t} + b_i (\mathbf{x}) \frac{\partial}{\partial x_i}
+ D_{ij} (\mathbf{x}) \frac{\partial^2}{\partial x_i \partial x_j}.
\label{eqn:L_SDE}
\eeq 
$D_{ij}$ is defined in Eqs.~\eqref{eqn:FP_operator_form_general} and repeated indices are summed over.

Given a function $v(t,x)$, the stochastic process $V_t = v(t, X_t)$  satisfies the SDE
\beq 
\dd V_t = \big[ L v(t, X_t ) \big] \dd t + \sum_{i=1}^n \sum_{\alpha=1}^m  
 \big[ \partial_{x_i} v(t, X_t) \big] \sigma_{i\alpha}(t, X_t) \ \dd W_\alpha (t) 
\label{eqn:dV_t_general}
\eeq  

If one can construct a Lyapunov function $v(t,x)$ that is positive define ''in Luapunov sense`` (i.e., $v(t,x)>0$ for $x \neq x_0$ and $v(t,x_0)=0$)  and at the same time
\beq 
\mathbb{E}[ \dd V_t ] \leq 0 \Leftrightarrow L v(t,x) \leq 0 \text{ for all } t\geq 0, 
\label{eqn:stable_req_01}
\eeq  
then $x_0$ is a stable in probability, i.e., 
\beq 
\lim_{x \rightarrow x_0} \mathbf{P} \big\{ \sup_{t>s} | X^{s,x}(t) - x_0 | > \varepsilon \big\} = 0.
\eeq 
[Chapter 11.2 in Ref.~\cite{Book_SP_Arnold} and Theorem 5.3 of Ref.~\cite{Book_SP_Khasminskii}.] Note that $v(t,x)$ is a deterministic scalar function and the most crucial step in this analysis is to identify the proper Lyapunov function $v(t,x)$.

For $B=0$, we identify the Lyapunov function $v(t,\theta) = v(\theta)$ as
\beq 
\begin{aligned}
& v (\theta) = |\cos \theta|, \text{ where } v (\theta_0) = 0 \text{ for } \theta_0 = \pm \frac{\pi}{2} + 2 N \pi.  \\
%& v (\theta) = \begin{cases} 
%    \cos \theta & \theta \in \big[ \frac{-\pi}{2}, + \frac{\pi}{2} \big]  \\
%    -\cos \theta & \theta \in \big[ \frac{\pi}{2}, + \frac{3\pi}{2} \big] 
%    \end{cases}  \\
%%%%%%%%%%%%%%
& \mathbb{E}[ \dd V ] =  \mathbb{E}\big[ v_{\theta} \dd \theta + \frac{1}{2} v_{\theta \theta} (\dd \theta)^2 \big] = -\frac{1}{2} | \cos \theta| \leq 0. %\\
%%%%%%%%%%%
%& \text{Conclusion: $\theta_0 $ is stable.} 
%& \text{i.e., does not necessarily go to 0 asymptotically. } 
\end{aligned}
\label{eqn:B0_Lyapunov_2}
\eeq 
$\theta_0 $ is the stable solution.

\section{SDE for likelihood function and its derivatives} \label{app:SDE_LL}

We derive the stochastic differential equation for $\frac{\partial}{\partial B} \log L_t \equiv l^B_t$ and $\frac{\partial^2}{\partial B^2} \log L_t \equiv l^{BB}_t$, the first and second derivatives  of the log-likelihood function with respect to $B$. The former is used in estimating $B$ on-the-fly.
To compute $l_B$, we first explicitly write [recall $L_t = \text{tr} [\tilde{\rho}_t]$]
\begin{subequations}
\begin{align}
l^B_t &= \frac{\partial \log L_t}{\partial B} = \frac{\partial_B L_t}{L_t} \equiv \frac{L^B_t}{L_t} ,
\label{eqn:lB_t_def} \\
%%%%%%%%%
\dd l^B_t &= \dd \big(\frac{1}{L_t}\big) \cdot L^B_t + \frac{1}{L_t} \cdot \dd L^B_t + \dd \big(\frac{1}{L_t}\big) \cdot \dd L^B_t \label{eqn:dlB_t_def}
%%%%%%%%%%%%%
%l^{BB}_t &= \frac{\partial^2 \log L_t}{\partial B^2} = \frac{\partial}{\partial B} \big[ \frac{L_t^B}{L_t} \big]  = \frac{L^{BB}_t}{L_t} - \big( \frac{L^{B}_t}{L_t} \big)^2  = \frac{L^{BB}_t}{L_t}  - \big( l_t^B \big)^2,  \label{eqn:lBB_t_def} \\
%%%%%%%%%
%\dd l^{BB}_t &= \dd \big(\frac{1}{L_t}\big) \cdot L^{BB}_t + \frac{1}{L_t} \cdot \dd L^{BB}_t + \dd \big(\frac{1}{L_t}\big) \cdot \dd L^{BB}_t -  \big[ 2 l^B_t \, \dd l^B_t + (\dd l^B_t)^2 \big] \label{eqn:dlBB_t_def}
\end{align}
\end{subequations}
Our goal is to replace the right-hand side of \eqref{eqn:dlB_t_def}% and \eqref{eqn:dlBB_t_def}
by $\rho_t$,  its derivatives, and measurement outcome $\dd Y_t$. Following the definition
\beq
\begin{aligned}
\dd L_t &= L_t \, \text{tr}[ \mathcal{M}(\rho_t) ] \, \dd Y_t \\
%%%%%%%%%5
\dd \left[ \frac{1}{L_t}  \right] &=  \frac{ -\text{tr}[\mathcal{M}(\rho_t)] }{L_t}
\left( \dd Y_t - \text{tr}[\mathcal{M}(\rho_t)] \dd t \right) \\
%%%%%%%%%%%%
\dd (\partial_B L_t) &\equiv \dd L^B_t= L^B_{t+dt}-L^B_t = L_t \, \text{tr}[ \mathcal{M}(\tau_t) ] \, \dd Y_t %\\
%%%%%%%
%\dd (\partial^2_B L_t) &\equiv \dd L^{BB}_t= L^{BB}_{t+dt}-L^{BB}_t = L_t \, \text{tr}[ \mathcal{M}(\eta_t) ] \, \dd Y_t
\end{aligned}
\eeq
where $\mathcal{M}(\rho_t)$ is defined in Eq.~\eqref{eqn:diffusion_DM_unnormalized} and
\beq
\begin{aligned}
\tau_t &\equiv \frac{ \tilde{\rho}^B_t }{ \text{tr}[ \tilde{\rho}_t ] } = \frac{ \tilde{\rho}^B_t }{ L_t } \Rightarrow \text{tr}[\tau_t] = \frac{L_t^B}{L_t} = \partial_B \log L_t = l_t^B.
%\\
%\eta_t &\equiv \frac{ \tilde{\rho}^{BB}_t }{ \text{tr}[ \tilde{\rho}_t ] } = \frac{ \tilde{\rho}^{BB}_t }{ L_t } \Rightarrow \text{tr}[\eta_t] = \frac{L_t^{BB}}{L_t}.
\end{aligned}
\eeq
%With these quantities Eq.~\eqref{eqn:lBB_t_def} is
%\beq
%l^{BB}_t = \frac{L^{BB}_t}{L_t}  - \big( l_t^B \big)^2 = \text{tr}[\eta_t]  - \big( \text{tr}[\tau_t] \big)^2 \label{eqn:lBB_t_def_tau_eta}
%\eeq

The SDE can be  derived from the Kraus operators $\Omega$ and $\partial_B \Omega \equiv \Omega_B$.
\beq
\begin{aligned}
\tilde{\rho}_{t+dt} &= \Omega \tilde{\rho}_{t}  \Omega^\dagger
\Rightarrow L_{t+dt} = \text{tr}\big[ \tilde{\rho}_{t+dt} \big]
= L_t \cdot \text{tr}\big[ \Omega {\rho}_{t}  \Omega^\dagger \big]  , \\
%%%%%%%%%%%%%%
\tilde{\rho}^B_{t+dt} &= \Omega_B \tilde{\rho}_{t}  \Omega^\dagger
+ \Omega \tilde{\rho}_{t}  \Omega_B^\dagger +
\Omega \tilde{\rho}^B_{t}  \Omega^\dagger. %, \\
%%%%%%%%%%%
%\tilde{\rho}^{BB}_{t+dt} &= 2\big( \Omega_B \tilde{\rho}^B_{t}  \Omega^\dagger
%+ \Omega \tilde{\rho}^B_{t}  \Omega_B^\dagger + \Omega_B \tilde{\rho}_{t}  \Omega_B^\dagger \big)+ \Omega \tilde{\rho}^{BB}_{t}  \Omega^\dagger. % \\
%%%%%%%%%%%
%\tilde{\rho}^{BBB}_{t+dt} &= 6\ \Omega_B \tilde{\rho}^B_{t}  \Omega_B^\dagger
%+ 3 \big( \Omega \tilde{\rho}^{BB}_{t}  \Omega_B^\dagger + \Omega_B \tilde{\rho}^{BB}_{t}  \Omega^\dagger \big)+
%\Omega \tilde{\rho}^{BBB}_{t}  \Omega^\dagger
\end{aligned}
\eeq
Normalizing by the likelihood function using the first equation, we get
\beq
\begin{aligned}
\tau_{t+dt} &= \frac{ \tilde{\rho}^B_{t+dt} }{L_{t+dt} } = \frac{1}{ \text{tr}\big[ \Omega {\rho}_{t}  \Omega^\dagger \big]  } \bigg[\Omega_B {\rho}_{t}  \Omega^\dagger
+ \Omega {\rho}_{t}  \Omega_B^\dagger +
\Omega \tau_{t}  \Omega^\dagger \bigg]. %, \\
%%%%%%%%%%%%%%%
%\eta_{t+dt} &= \frac{ \tilde{\rho}^{BB}_{t+dt} }{L_{t+dt} } = \frac{1}{ \text{tr}\big[ \Omega {\rho}_{t}  \Omega^\dagger \big]  } \bigg[ 2\big( \Omega_B \tau_{t}  \Omega^\dagger
%+ \Omega\tau_{t}  \Omega_B^\dagger + \Omega_B {\rho}_{t}  \Omega_B^\dagger \big)+ \Omega \eta_{t}  \Omega^\dagger \bigg]. %, \\
%%%%%%%%%%%%%%%
%\xi_{t+dt} &= \frac{ \tilde{\rho}^{BBB}_{t+dt} }{L_{t+dt} } = \frac{1}{ \text{tr}\big[ \Omega {\rho}_{t}  \Omega^\dagger \big]  } \bigg[ 6\ \Omega_B \tau_{t}  \Omega_B^\dagger
%+ 3 \big( \Omega \eta_{t}  \Omega_B^\dagger + \Omega_B \eta_{t}  \Omega^\dagger \big)+
%\Omega \xi_{t}  \Omega^\dagger \bigg].
\end{aligned}
\eeq
Note that $\rho_t$ and $\tau_t$ %, $\eta_t$,
are properly normalized at each time $t$ so their values remain finite in long $t$ limit. $\tilde{\rho}_t$ becomes exponentially small and so are its $B$-derivatives. This evolution is numerically stable in terms of evolution \cite{Albarelli2018restoringheisenberg}.

Now we consider $\dd l^B_t$ defined in \eqref{eqn:dlB_t_def}.  %and \eqref{eqn:dlBB_t_def}
Using $\dd Y_t^2=\dd t$, we get
\beq
\begin{aligned}
\dd l^B_t &= l^B_{t+dt} - l^B_t= \underbrace{
\bigg(\text{tr}[\mathcal{M}(\tau_t)]  - \text{tr}[\mathcal{M}(\rho_t)] l^B_t \bigg) }_{= \text{tr} \big[ \mathcal{M}( \tau_t - \rho_t l_t^B ) \big] = \text{tr} \big[ \mathcal{M}( \rho_t^B ) \big] }
\cdot \bigg( \dd Y_t - \text{tr}[\mathcal{M}(\rho_t)]  \, \dd t \bigg) \\
%%%%%%%%%
&=
\bigg(\text{tr}[\mathcal{M}(\tau_t)]  - \text{tr}[\mathcal{M}(\rho_t)] \cdot \text{tr}[\tau_t] \bigg)
\cdot \bigg( \dd Y_t - \text{tr}[\mathcal{M}(\rho_t)]  \, \dd t \bigg)
%%%%%%%%%%%%%%%%
\end{aligned}
\label{eqn:dl^B_dt_SDE}
\eeq
We have used
\beq
\rho^B_t = \frac{\partial}{\partial B} \bigg[ \frac{\tilde{\rho}_t }{L_t} \bigg]
= \frac{\tilde{\rho}^B_t L_t - L_t^B \tilde{\rho}_t }{L_t^2}
= \tau_t - \rho_t l_t^B = \tau_t - \rho_t \text{tr}[\tau_t].
%\label{eqn:rho^B}
\nonumber
\eeq
$\dd l^B_t $ in Eq.~\eqref{eqn:dl^B_dt_SDE} is used for updating $B_\text{est}$ in the main text.

With Eq.~\eqref{eqn:dl^B_dt_SDE}, $l^B_t = \sum_{t'<t} \dd l^B_{t'}$ with the initial condition $l^B_{t=0} =0$. $l^B_t$ can also be evaluated using finite-difference:
\beq
l^B_t(B_0) \approx \frac{l_t(B_0 + \dd B/2)- l_t(B_0 - \dd B/2)}{\dd B}.
\label{eqn:dl^B_dt_FD}
\eeq
Eq.~\eqref{eqn:dl^B_dt_FD} is used to verify Eq.~\eqref{eqn:dl^B_dt_SDE}; the good agreement is seen in Fig.~\ref{fig:online_parameter_est}(b).

%For completeness we give the following expressions:
%\beq
%\begin{aligned}
%\tau^B_t &= \frac{\partial}{\partial B} \bigg[ \frac{\tilde{\rho}^B_t }{L_t} \bigg]
%= \frac{\tilde{\rho}^{BB}_t L_t - L_t^B \tilde{\rho}^B_t }{L_t^2}
%= \eta_t - \tau_t l_t^B = \eta_t - \tau_t \text{tr}[\tau_t] \\
%\Rightarrow \ & \text{tr}[ \tau^B_t ] = \text{tr}[\eta_t] - \big( \text{tr}[\tau_t]  \big)^2. \\
%%%%%%%%%%%%%
%\eta^B_t &= \frac{\partial}{\partial B} \bigg[ \frac{\tilde{\rho}^{BB}_t }{L_t} \bigg]
%= \frac{\tilde{\rho}^{BBB}_t L_t - L_t^B \tilde{\rho}^{BB}_t }{L_t^2}
%= \xi_t - \eta_t l_t^B = \xi_t - \eta_t \text{tr}[\tau_t] \\
%\Rightarrow \ & \text{tr}[ \eta^B_t ] = \text{tr}[\xi_t] - \text{tr}[\eta_t] \cdot \text{tr}[\tau_t] .
%\end{aligned}
%\label{eqn:tau^B}
%\eeq
%and by partial differentiating \eqref{eqn:rho^B} with respect to $B$
%\beq
%\begin{aligned}
%\rho^{BB}_t &= \frac{\partial}{\partial B} \rho^B_t
%= \tau^B_t - \rho^B_t \text{tr}[\tau_t] - \rho_t \text{tr}[\tau^B_t] \\
%%%%%%%%%%
%&= \underbrace{ \big( \eta_t - \tau_t \text{tr}[\tau_t] \big)  }_{ \tau^B_t } - \underbrace{  \big( \tau_t - \rho_t \text{tr}[\tau_t] \big) }_{ = \rho^B_t} \text{tr}[\tau_t]
%- \rho_t \underbrace{ \big( \text{tr}[\eta_t] - \big( \text{tr}[\tau_t]  \big)^2 \big) }_{ = \text{tr}[ \tau^B_t ] } \\
%%%%%%%%%%%%%%%%%
%&= \eta_t - \tau_t \text{tr}[\tau_t]  + \rho_t \big( 2 \big( \text{tr}[\tau_t]  \big)^2 - \text{tr}[\eta_t] \big).
%%%%%%%%%%%%%
%\end{aligned}
%\label{eqn:tau^BB}
%\eeq

%%%%%%%%%%%%%%%%%%%%%%%%%%%%%%%%%%%%%%%%%%%
\bibliography{parameter_estimation}
%\bibliographystyle{unsrt}
%\begin{thebibliography}{10}
%\end{thebibliography}

%\appendix

\end{document}